\documentclass[aps,prb,twocolumn,showpacs]{revtex4}

\usepackage{latexsym}
\usepackage{graphicx}
\usepackage{times,psfrag,subfigure}
\usepackage{amsmath}
\usepackage{dcolumn}
\usepackage{latexsym,amsmath,amssymb,bm,euscript}
\usepackage{afterpage}
\newcommand{\bra}[1]{\ensuremath{\langle#1|}}
\newcommand{\ket}[1]{\ensuremath{|#1\rangle}}

\newcommand{\NiGaS}{NiGa$_2$S$_4$}
\newcommand{\nnij}{\ensuremath{\langle ij \rangle}}
\newcommand{\avg}[1]{\ensuremath{\langle #1 \rangle}}

\begin{document}

\title{Quadrupolar correlations and spin freezing in $S=1$ triangular
  lattice antiferromagnets} 
\date{\today}
\author{E.M.~Stoudenmire}
\affiliation{Department of Physics, University of California, Santa Barbara, CA 93106}
\author{Simon Trebst}
\affiliation{Microsoft Research, Station Q,  University of California, Santa Barbara, CA 93106}
\author{Leon Balents}
\affiliation{Kavli Institute for Theoretical Physics, Santa Barbara, CA 93106}

\begin{abstract}
  Motivated by experiments on \NiGaS, we discuss characteristic (finite
  temperature) properties of spin $S = 1$ quantum antiferromagnets on
  the triangular lattice.  Several recent theoretical studies have
  suggested the possibility of quadrupolar (spin-nematic) ground states
  in the presence of sufficient biquadratic exchange.  We argue that
  quadrupolar {\sl correlations} are substantially more robust than the
  spin-nematic ground state, and give rise to a two peak structure of
  the specific heat. We characterize this behavior by a novel $T > 0$
  semiclassical approximation, which is amenable to efficient Monte
  Carlo simulations.  Turning to low temperatures, we consider the
  effects of weak disorder on incommensurate magnetic order, which
  is present when interactions beyond nearest neighbor exchange are
  substantial.  We show that non-magnetic impurities act as random
  fields on a component of the order parameter, leading to the
  disruption of long-range magnetic order even when the defects are
  arbitrarily weak. Instead, a gradual freezing phenomena is expected on
  lowering the temperature, with no sharp transition but a rapid slowing
  of dynamics and the development of substantial spin-glass-like
  correlations. We discuss these observations in relation to
  measurements of \NiGaS.
\end{abstract}


\maketitle


\section{Introduction}

The triangular lattice antiferromagnet has long been of interest because
of its potential to exhibit exotic phases as a result of frustration and
because it is the underlying lattice of many real materials.
Unfortunately though, the simplest triangular lattice models have a well
known tendency to order into states that are basically classical. For
instance, it is generally believed that the nearest-neighbor spin 1/2
triangular lattice antiferromagnet orders into the 120-degree spiral
state at zero temperature. However, since the models that describe real
triangular lattice systems generally contain additional interactions,
the possibilities for more interesting behavior remain vast.

One triangular lattice system that was recently proposed as a candidate
for exotic behavior was the spin one antiferromagnet \NiGaS, which
exhibits no long-range spin ordering at zero temperature despite a low
temperature susceptibility and specific heat\cite{nakatsuji-05}  that suggest
two-dimensional magnons; in particular, the specific heat grows very
clearly as $T^2$ below $10K$. The specific heat furthermore shows no
signs of a phase transition but instead shows two rounded peaks at $10K$
and $80K$, the latter being the Curie-Weiss temperature scale. Neutron
scattering reveals the presence of short range magnetic ordering below
$50K$ with an incommensurate wavevector; however, the magnetic
correlation length never grows past about 7 lattice spacings even as the
system is cooled to $1.5K$.\cite{nakatsuji-05,nakatsuji-07} More recent
studies have shown that local moments do form and {\sl gradually} freeze
at low temperatures, fluctuating increasingly slowly as the
temperature is reduced below
$10K$.\cite{takeya2008sda,yaouanc:092403,citeulike:2437587} They
eventually become static on the scale of the slowest available
experimental local probe (muon spin resonance) below about~$2K$.

A series of theoretical studies of this system subsequently appeared
which attributed the thermodynamic signatures of ordering to a
quadrupolar (also called spin-nematic) ground state stabilized by a
nearest-neighbor biquadratic coupling term in the
Hamiltonian.~\cite{tsunetsugu-06,laeuchli-06,senthil-06} Though such a
state can account for the observed low temperature thermodynamics
without requiring there to be any long-range spin ordering, it is
inconsistent with the observed short-range incommensurate spiral order
at the lowest temperatures.  Moreover, the observation of static (though
spatially random) spins at low temperatures casts doubt on the relevance
of the spin-nematic phase to \NiGaS.  Indeed, to date no clear specific
signature of even quadrupolar correlations (much less order) has been
identified in \NiGaS.

In this paper, we reconsider the principal experimental observations,
and argue that such correlations are likely present in \NiGaS, and
signified by the unusual two peak structure in the specific heat.  More
formally, the double peak is a signature of proximity to a zero
temperature quantum critical point between an antiferromagnetically
ordered and a spin-nematic state.  We also suggest a mechanism for the
low temperature spin freezing, which  is {\sl independent} of the
spin-nematic correlations.  This mechanism should apply to {\sl any}
strongly two-dimensional magnet whose ground state in the absence of
disorder is an incommensurate spiral.  We argue that both these
phenomena (the two peak specific heat and spin freezing) can be
understood on purely symmetry grounds.  Thus they may be expected to
occur much more generally in other materials under conditions that we
discuss.

For concreteness, we work within an extended model that not only
includes the nearest-neighbor Heisenberg and biquadratic couplings, but
also third-nearest-neighbor Heisenberg interactions:
\begin{equation}
H = J_3 \sum_{[ij]} {\bf S}_i\cdot{\bf S}_j + J_1 \sum_{\nnij}
{\bf S}_i\cdot{\bf S}_j - K \sum_{\nnij}
\left({\bf S}_i\cdot{\bf S}_j\right)^2.\label{eq:1}
\end{equation}
The third neighbor couplings are denoted by $[ij]$ (on the
triangular lattice, the third neighbor bonds point along the first neighbor
bonds but  doubled in length).  This additional $J_3$ term is perhaps the
simplest way to account for the observed order, which is described by a
wavevector that is slightly less than half of the 120-degree state
wavevector. Indeed, such an incommensurate spiral state is the exact
solution of the classical limit of the above model in the limit $J_3 > -
J_1 > K \geq 0$. Furthermore, detailed microscopic studies of \NiGaS\ indicate that
a strong third-neighbor interaction is in fact present while
second-neighbor interactions appear to be much less important.~\cite{mazin-07,takubo:037203}

Let us outline our approach to the problem, and the layout of the
remainder of the paper.  For $J_3=0$, Eq.~(\ref{eq:1}) has been analyzed
already in Refs.~\onlinecite{tsunetsugu-06,laeuchli-06}, and shown to
exhibit a non-magnetic but (ferro-)quadrupolar (spin-nematic) ground
state in a broad parameter range.  We argue that this remains true for
the more general model with $J_3\neq 0$ in Sec.~\ref{sec:J3-model}.  In addition, when
$K$ is not too large, Eq.~(\ref{eq:1}) has magnetically ordered ground
states.  Each of these states is connected to the quadrupolar one by a
quantum phase transition (QPT).  A major portion of this paper is
concerned with establishing the $T>0$ properties of systems in the
vicinity of such a QPT.  We first present general symmetry-based
arguments for the existence of two energy scales, and a resulting two peak
structure of the specific heat, in Sec.~\ref{sec:symm-order-param}.  To verify these ideas, we
then introduce a novel ``semiclassical SU(3)'' or ``sSU(3)" approximation,
in which we can obtain these properties for Eq.~(\ref{eq:1}) by
classical Monte Carlo simulations.  Though this approach is not expected
to be quantitatively accurate for the $S=1$ model of interest, it does
preserve all the symmetries of Eq.~(\ref{eq:1}) and a qualitatively appropriate
$T=0$ limit, unlike conventional semiclassical approaches.  We carried out
extensive numerical simulations using the sSU(3) approach to establish the
nature of the $T>0$ phase transitions and crossovers, which we
describe in Secs.~\ref{sec:nn-model} and \ref{sec:J3-model}. 

Next, we turn to the effects of disorder, assuming that the putative
ground state in the absence of impurities is magnetically ordered
(though it may be near the QPT separating it from the quadrupolar
phase).  We show that the generic {\sl incommensurate spiral} ground
state of Eq.~(\ref{eq:1}) is highly susceptible to defects.  This is a
direct consequence of the nature of the associated order parameter,
which leads to strong ``random field'' effects of impurities -- effects
which do not pertain to the more conventional 120$^\circ$
three-sublattice ground state.  Moreover, very general arguments due to
Imry and Ma show that such random field systems {\sl in two dimensions}
-- but not in three -- are always disordered.  Hence we are led to
conclude that the nature of the order parameter (which in turn is due to
the large $J_3$ interaction) and the very weak three-dimensional
coupling in NiGa$_2$S$_4$ are responsible for the observed low
temperature glassy state.  This mechanism for spin freezing is discussed
in Sec.~\ref{sec:corr-long-scal}.  We conclude the paper with a
discussion in Sec.~\ref{sec:discussion}.  The Appendix describes details
of the measurements used in the Monte Carlo simulations.


\section{Symmetry and order parameters}
\label{sec:symm-order-param}

In this section, we discuss the symmetries and order parameters of the
Hamiltonian in Eq.~\eqref{eq:1} and its various proposed ground states.
The Hamiltonian itself has global SU(2) spin-rotation symmetry, and the
full space group symmetry of the triangular lattice.  It is also time
reversal invariant.  A subset of these symmetries are retained in the
various ground states.  First, we discuss the conventional magnetically
ordered states.  We consider only coplanar spirals, which dominate the
classical ground states of Eq.~\eqref{eq:1}.  In these states, the
pattern of spin ordering is described by
\begin{equation}
  \label{eq:2}
  \langle {\bf S}_i \rangle = {\rm Re}\, \left[ {\bf d} e^{i {\bf
        k}\cdot {\bf r}_i} \right],
\end{equation}
where ${\bf d}= |d|({\bf\hat e}_1 + i {\bf \hat e}_2)$, with ${\bf\hat
  e}_1, {\bf\hat e}_2$ orthogonal unit vectors, and ${\bf k}$ is the
spiral wavevector.  The spins lie in the plane spanned by ${\bf\hat
  e}_1, {\bf\hat e}_2$, normal to ${\bf \hat e}_3 = {\bf \hat e}_1
\times {\bf \hat e}_2$.   Clearly, the spin spiral breaks spin
rotational symmetry, and, assuming ${\bf k}\neq 0$, it does so
completely: rotation by any angle about any spin axis alters the spin
state.  Moreover, under the same assumption, the spiral breaks
translational symmetry.  However, an important symmetry is retained by
the coplanar spiral.  In particular, under the combined action of a
translation by any (lattice) vector ${\bf a}$, {\sl and} a spin rotation
by the angle ${\bf k}\cdot {\bf a}$ about the ${\bf\hat e}_3$ spin axis,
the state is unchanged.  The situation for spatial rotations is more
complex.  Consider the three-fold ($C_3$) rotation about a lattice site, a
symmetry of the triangular lattice.  In the familiar $120^\circ$ state
(the ground state for $K=J_3=0$), the ordered spins remain invariant
under this rotation, since the sublattices are unchanged.  This
invariance is, however, not generic.  Formally, it can be traced to the
fact that in this case, ${\bf k}$ is a zone-corner wavevector (the ``K
point''), which is invariant under such a rotation.  Any other ${\bf k}$
is {\sl not} invariant under a three-fold rotation, and moreover, the
invariance cannot be regained by combining the spatial rotation with an
SU(2) one.  The breaking of the discrete $C_3$ rotation symmetry of $H$
implies that there are then three distinct ``domains'' of spiral state,
with each of the three possible wavevectors forming an invariant set
under the rotations.  Each domain cannot be continuously transformed into
another.  This is quite different from the $120^\circ$ state, in which
all possible spin configurations can be smoothly transformed into one
another.   

Next consider the (ferro)quadrupolar ground states.  In them the ordered
spin moment is zero:
\begin{equation}
  \label{eq:3}
  \langle {\bf S}_i \rangle_{Q} = 0.
\end{equation}
However, the average magnetic quadrupole moment is non-zero:
\begin{eqnarray}
  \label{eq:4}
  Q_{i;\mu\nu} & =  & \frac{1}{2}\langle \left\{ S_i^\mu, S_i^\nu\right\} \rangle -
  \frac{2}{3} \delta_{\mu\nu} , \\
  Q_{\mu\nu} & = & \frac{1}{N} \sum_i Q_{i;\mu\nu},
\end{eqnarray}
where $Q_{\mu\nu}$ is the (ferro-)quadrupolar order parameter.  By construction,
$Q$ is a traceless symmetric matrix.  In the ``spin-nematic'' states
which we consider, the eigenvalues of this matrix come in only two
values, i.e. ${\rm eigs}(Q) = \{q,-q/2,-q/2\}$.  More complicated ``biaxial
spin-nematic'' states are possible in principle, in which all three
eigenvalues are distinct.  We will not consider them here, as there is
no obvious reason for them to occur in Eq.~\eqref{eq:1}.  In the
remainder of the paper, when we refer to the quadrupolar state, it will
always be assumed to be of the (single axis) spin-nematic type.  Then a
general $Q$ can be written as
\begin{equation}
  \label{eq:5}
  Q_{\mu \nu} = \frac{3}{2} q\left( \hat{n}_\mu \hat{n}_\nu - \frac{1}{3}
    \delta_{\mu\nu}\right),
\end{equation}
where ${\bf\hat n}$ is a unit vector known as the ``director''.  The
change of the director by a sign, ${\bf\hat n} \rightarrow - {\bf\hat
  n}$, has no physical significance, because $Q$ is even in ${\bf\hat
  n}$.  The quadrupolar state breaks spin rotation symmetry, but retains
invariance under spin rotations about the axis of the director.
Furthermore, it retains full space group and time reversal symmetry.  

Comparison of the symmetries of the magnetically ordered and quadrupolar
ground states reveals an important fact: the symmetry group of the
spiral state is a {\sl subgroup} of the symmetry group of the
quadrupolar one.  Thus any order parameter of the quadrupolar state must
already be non-zero in the magnetic phase.  Indeed, within Landau
theory, one expects a coupling between the magnetic and quadrupolar
order parameters, i.e. a term in the free energy density $f$ of the form
\begin{equation}
  \label{eq:6}
  f_{Q-d} = \lambda Q_{\mu\nu}\left[ d_\mu^* d_\nu^{\vphantom*}+d_\nu^*
    d_\mu^{\vphantom*} - \tfrac{2}{3} {\bf d}^* \cdot {\bf d}^{\vphantom*}\delta_{\mu\nu}\right],
\end{equation}
where $\lambda$ is a non-zero coupling constant.  This generically
induces a non-zero $Q_{\mu\nu}$ of the form in Eq.~(\ref{eq:5}), with
$|q| \propto |d|^2$ and ${\bf\hat n} = {\bf\hat e}_3$.  

\begin{figure}[t]
\includegraphics[width=\columnwidth]{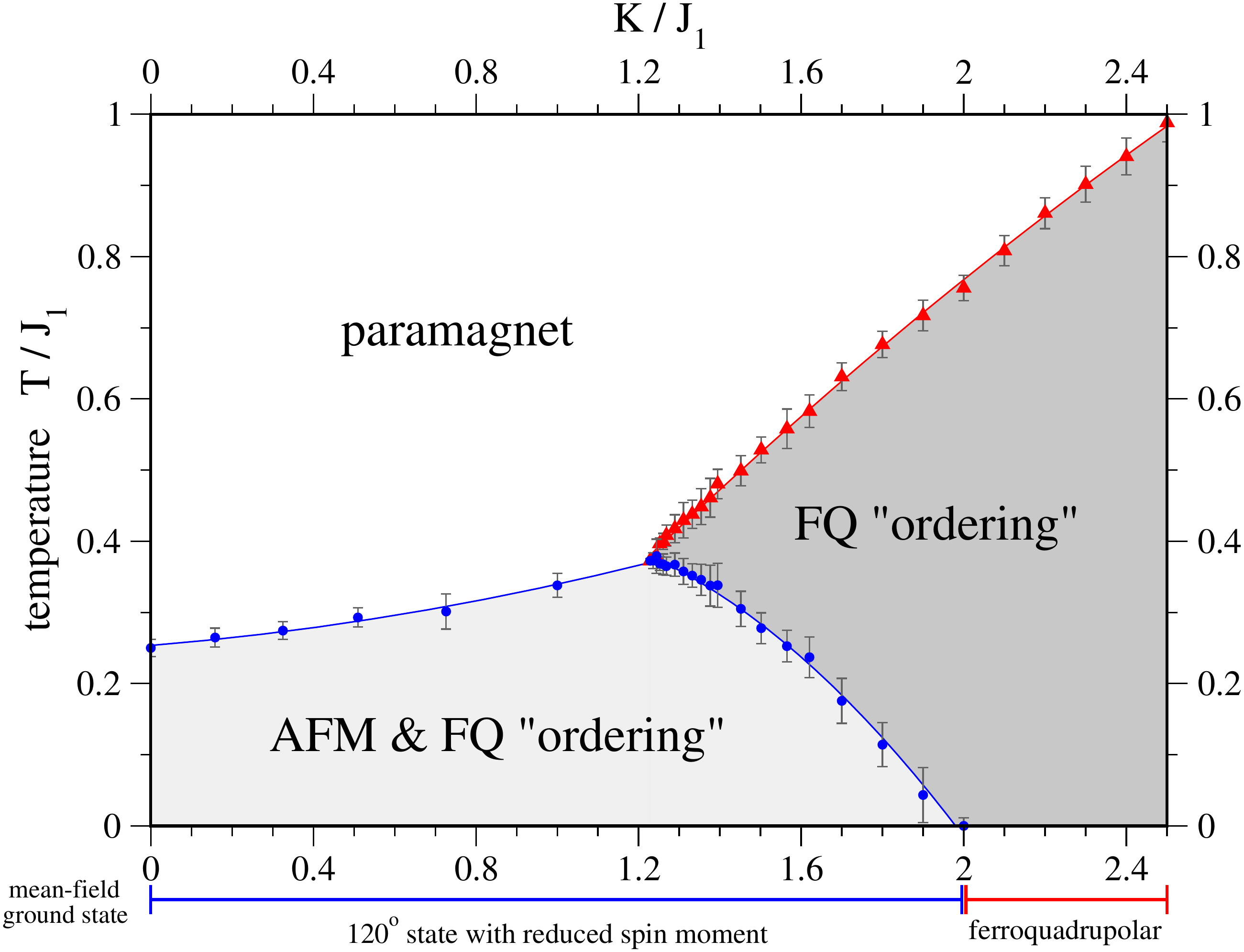}
\caption{
  Finite temperature phase diagram of the nearest-neighbor only model with  
  antiferromagnetic $J_1>0$  and a ferronematic $K>0$ obtained from
  numerical simulations in the sSU(3) approximation.
  The location of the phase boundaries are obtained from the position of the 
  specific heat peaks in our numerical simulations. Approaching zero temperature
  the semiclassical sSU(3) approximation recovers the ground states obtained
  in mean-field theory.
  We note that the phase boundaries indicate a rapid growth of the ferroquadrupolar 
  and magnetic correlation lengths, but not true long-range order excluded by the
  Mermin-Wagner theorem.
}
\label{fig:J1_K_two_peak}
\end{figure}

It follows that $q$ in Eq.~(\ref{eq:5}) remains non-zero across the QPT
between the magnetic and quadrupolar states.  Conversely, at this QPT,
presuming it is continuous or weakly first order, the magnetic order
parameter ${\bf d}$ becomes arbitrarily small.  Associated with each of
these order parameters is an energy scale, describing the energy needed
to disturb the order parameter locally.  This energy is supplied with
increasing temperature, releasing entropy associated with the ordering.
As usual, the specific heat is expected to show a peak at this
characteristic temperature (whose precise nature requires more detailed
considerations of symmetry, universality, and energetics).  Near the
QPT, on the magnetically ordered side, one is then led to expect two
widely separated energy scales: the quadrupolar ``ordering'' scale,
which remains of $O(1)$ (i.e. not parametrically small near the QPT),
and a much smaller magnetic ``ordering'' scale, which vanishes on
approaching the QPT.  There will be a specific heat peak at each of
these temperatures.  The schematic phase diagram is shown in
Fig.~\ref{fig:J1_K_two_peak}.  We have used ``ordering'' in quotes,
because in a two-dimensional system, the continuous order parameters are
prohibited from exhibiting {\sl long-range} order by the Mermin-Wagner
theorem.  However, the specific heat is dominated by short distance
correlations, and exhibits peaks regardless of the presence or absence
of true phase transitions.

\section{Semiclassical SU(3) Approximation}
\label{sec:semiclassical}

In the previous section, we argued for the emergence of two temperature
scales and a two peak structure in the specific heat, whenever the
system is in vicinity of the QPT between an antiferromagnetic and
quadrupolar ground state.  In what follows, we will check these
arguments by explicit calculations of $T>0$ properties.  To do this
presents a serious difficulty: theoretical approaches to calculate the
low but nonzero temperature behavior of frustrated quantum Hamiltonians
are extremely limited.  A direct attack with quantum Monte Carlo is
prevented by the sign problem, and many other approaches (e.g. variational
wavefunctions, Lanczos diagonalization) are restricted to ground state
properties.  

One natural approach is the classical approximation, in which the
quantum spin operators are replaced by continuous classical vectors of
length $S$.  For many quantum spin models, the success of spin-wave
theory, even for low spin, proves the applicability of this method.
However, for the study of Eq.~(\ref{eq:1}), such an approach is fatally
flawed.  The difficulty lies in the highly quantum nature of the
quadrupolar spin-nematic correlations induced by the biquadratic
interaction for spin $S=1$.  In particular, the ferronematic ground
states favored by $K>0$ are of ``easy-plane'' type, in which the spins
may be thought of as fluctuating quantum mechanically in the plane {\sl
  perpendicular} to the director $\hat{n}$ in Eq.~(\ref{eq:5}).  This
corresponds to the sign $q<0$. For large spin, including the classical limit,
however, the biquadratic interaction with $K>0$ favors collinear states
of the spins, and hence ``easy axis'' rather than easy-plane
ferronematic order (with the opposite sign of $q$).  In a simple
mean-field theory, the nature of the quadrupolar order changes
from easy plane when $S\leq 2$ to easy axis for $S>2$.  Therefore even
qualitatively the classical vector spin approximation fails to capture
the proper ferronematic correlations of the physics $S=1$ system.  

We thus follow an alternative approach, which generalizes the zero
temperature variational approach of Ref.~\onlinecite{laeuchli-06} to $T>0$.  In
that work, a trial ground state was taken to be entanglement-free,
i.e. of direct product form:
\begin{equation}
  \label{eq:12}
  |\Psi\rangle = \otimes_i |\psi_i\rangle,
\end{equation}
with an arbitary spin $S=1$ state $|\psi\rangle_i$ on site $i$.  A
general single spin state can be written as 
\begin{equation}
  \label{eq:13}
\ket{\psi_i} = b_{i1} \ket{x} + b_{i2} \ket{y} + b_{i3} \ket{z} \ ,
\end{equation}
with an arbitary complex vector ${\bf b}_i$ satisfying the normalization
constraint that ${\bf b}^{*}_i\cdot {\bf b}_i = 1$.  Here we have used the
time-reversal invariant basis of $S=1$ states defined by
\begin{eqnarray*}
\ket{x} & = & i\left(\frac{1}{\sqrt{2}} \ket{1} - \frac{1}{\sqrt{2}} \ket{\bar{1}} \right) \\
\ket{y} & = & \frac{1}{\sqrt{2}} \ket{1} + \frac{1}{\sqrt{2}} \ket{\bar{1}} \\
\ket{z} & = & -i \ket{0}
\end{eqnarray*}
where $\ket{1}$,$\ket{0}$ and $\ket{\bar{1}}$ are the usual $S=1$ states
quantized along the z-axis.  These basis states are null vectors of
their respective spin components, i.e. $S^x \ket{x}=S^y\ket{y}=S^z
\ket{z}=0$.  Semiclassically, the state
$\ket{x}$ may be viewed as one in which the 
spin fluctuates primarily in the yz-plane (and similarly for the other
two states).  

The expectation values of the spin and quadrupolar operators have simple
expressions in these states:
\begin{eqnarray}
  \label{eq:14}
  \bra{\psi_i} {\bf S}_i \ket{\psi_i} & = & -i {\bf b}_i^* \times {\bf
    b}_i^{\vphantom*}, \\ \nonumber
 Q_{i;\mu\nu} & = & \frac{1}{3} \delta^{\mu\nu} - \frac{1}{2}\left( b_{i\mu}^\dagger
    b_{i\nu}^{\vphantom\dagger} + b_{i\nu}^\dagger
    b_{i\mu}^{\vphantom\dagger} \right).
\end{eqnarray} 
Note that when  ${\bf b}_i$ is real (it may be multiplied by an overall
phase without affecting physical properties), the spin expectation vanishes.
Thus in this case it may be interpreted as the nematic
director, i.e. ${\bf b}_i = {\bf\hat{n}}_i$.   Adding an imaginary
part to ${\bf b}_i$ produces a nonzero spin moment that is orthogonal to
the plane in which the real and imaginary parts of ${\bf b}_i$ lie.

From Eq.~(\ref{eq:14}), we can compute the expectation value of the
Hamiltonian, 
\begin{eqnarray}
H_{\rm cl} & \equiv &  \bra{\Psi}H\ket{\Psi} \label{eqn:effective_h}  \\ 
& = & J_3 \sum_{[ij]} |{\bf b}^{*}_i\cdot{\bf b}_j|^2 - |{\bf
  b}_i\cdot{\bf b}_j|^2 \nonumber\\ 
 & & \mbox{} + \sum_{\nnij} J_1 |{\bf b}^{*}_i\cdot{\bf b}_j|^2 - (J_1 + K) |{\bf b}_i\cdot{\bf b}_j|^2 \ . \nonumber
\end{eqnarray}
In Ref.~\onlinecite{laeuchli-06}, this expression was minimized for the
nearest-neighbor model ($J_3=0$) to find variational ground states.  It
was found that the variational approximation qualitatively reproduced
the correct $T=0$ phase diagram, though it incorrectly determined the
quantitative location of the ferronematic-antiferromagnetic phase
boundary.  Note that, at $T=0$, the variational approximation is
equivalent to a mean-field theory in which the expectation values
$Q_{i;\mu\nu}$ and $\langle{\bf S}_i\rangle$ are self-consistently
determined from a Hamiltonian of decoupled $S=1$ spins.

To address $T>0$ temperature properties, we follow an unusual procedure
of elevating Eq.~(\ref{eqn:effective_h}) to a Hamiltonian of a
two-dimensional statistical mechanical model, allowing  the classical
${\bf b}_i$ vectors to fluctuate thermally with the Boltzmann weight
$\exp[-\beta H_{\rm cl}]$.  Because in this new classical problem, the
``spins'' appear as three-component complex vectors, and transform under
symmetries by rotations by unitary three-dimensional matrices, we call
this treatment the semiclassical SU(3) or sSU(3) approximation, and the
resulting statistical mechanical problem the sSU(3) model.  We
emphasize, however, that the model Eq.~(\ref{eqn:effective_h}) does not
have SU(3) symmetry, but only the physically appropriate SU(2) spin
rotational and lattice symmetries.  

The sSU(3) model can be understood as a leading-order term in a cumulant
expansion of the full quantum partition function.  Specifically, we
approximate
\begin{eqnarray}
  \label{eq:15}
  Z & = & {\rm Tr}\, e^{-\beta H} \\
  & = & \int \prod_i d\Omega_{b_i} \bra{\Psi(b)} e^{-\beta H} \ket{\Psi(b)}
  \nonumber \\
  & \approx & \int \prod_i d\Omega_{b_i} e^{-\beta \bra{\Psi(b)}H \ket{\Psi(b)}} =
  \int \prod_i d\Omega_{b_i} e^{-\beta H_{cl}}, \nonumber
\end{eqnarray}
where $d\Omega_b = (2\pi)^2 [\prod_{\mu=x,y,z} d{\rm Re} \,b_\mu d{\rm
  Im}\, b_\mu]\delta(\sum_\mu |b_\mu|^2 -1)$ is the appropriately
normalized measure over unit 3-component complex vectors, such that
$\int d\Omega_b 1 = {\rm Tr} 1 = 3$.  From this perspective, we see that
the sSU(3) approximation becomes exact at high temperatures (reproducing
the exact free energy to $O(\beta)$ in the $\beta$ expansion),
and approaches the variational result as $T\rightarrow 0$.  It can
therefore be expected to yield a reasonable qualitative approximation to
the physics over the full range of intermediate temperatures.

One may consider at this stage, given that the $T=0$ variational
approach is equivalent to mean-field theory, why not instead apply the
usual mean-field theory at $T>0$?  The reason to choose the sSU(3)
approach is that it captures much better the effects of thermal
fluctuations, which are very important in a two-dimensional system.
For instance, because Eq.~(\ref{eqn:effective_h}) has the physically correct
symmetries, we expect by universality that it displays the correct {\sl
  exact} critical behavior at any $T>0$ phase transitions.  Moreover, it
avoids spurious transitions which are in fact prohibited by the
Mermin-Wagner theorem.  In contrast, straight mean-field theory will
produce sharp phase transitions associated with quadrupolar {\sl and}
magnetic ordering, neither of which are in fact allowed in two
dimensions.  

Despite the above advantages of the sSU(3) approximation over mean-field
theory, it is very amenable to numerical computations.  The classical
Hamiltonian in Eq.~(\ref{eqn:effective_h}) is completely real, so there
is manifestly no sign problem.  The only subtlety lies in the gauge
redundancy of the ${\bf b}_i$ variables, whose phase (on each site) has
no physical significance.  However, no difficulty is incurred in Monte
Carlo simulations by simply including the redundancy, and restricted
measurements to gauge-invariant observables.  
For our numerical simulations of this semiclassical Hamiltonian we have 
built our implementation of a classical Monte Carlo algorithm on the ALPS 
libraries \cite{alps}, and in particular expanded its classical Monte Carlo 
application code to sample configurations of the three-component 
complex vectors or ``spins" as introduced above. 
We give a detailed account of how to measure various physical observables 
in this Monte Carlo scheme in the Appendix.

\section{Finite Temperature Properties of the Nearest-Neighbor Model}
\label{sec:nn-model}

As a first step we apply the semiclassical sSU(3) approximation to
numerically analyze the $T>0$ thermodynamic features and
correlations of the simplest, nearest-neighbor Hamiltonian with
biquadratic exchange
\begin{equation}
H = J_1 \sum_{\nnij} {\bf S}_i\cdot{\bf S}_j - K \sum_{\nnij} \left({\bf S}_i\cdot{\bf S}_j\right)^2 \,,
\label{eq:j1k-Hamiltonian}
\end{equation}
where we consider antiferromagnetic $J_1 > 0$ and ferronematic $K > 0$. 
Though our primary focus will be on finite temperature properties of this model,
it is helpful to begin by reviewing what is already known about its ground states.

\subsection{Ground State Properties}

The full zero temperature phase diagram of the nearest-neighbor
bilinear-biquadratic model contains the conventional ferro- and
antiferromagnetic magnetic phases when $K$ is small, but also includes a
ferroquadrupolar phase and an antiferroquadrupolar phase stablized by
larger values of $K$.

In the range of parameters of interest to us ($J_1$ and $K$ positive),
there is a single quantum phase transition as shown in
Fig.~\ref{fig:J1_K_two_peak}.  For small $K/J_1$, the ground state is a
120-degree spiral ground state with a reduced spin moment.  Then, as
$K/J_1$ is increased, the spin moment decreases to zero leaving a state
with only ferroquadrupolar order i.e.\ no average spin moment but an
identical director (identical, real ${\bf b}$) on every site. In the
mean-field/variational approximation, this phase transition happens
precisely at $K/J_1 = 2$.  Exact diagonalization results~\cite{laeuchli-06}
for the quantum model show, however, that quantum fluctuations strongly 
reduce this value to about $K/J_1 = 0.4$.

Within mean-field theory, a detailed study shows that the quadrupolar to
antiferromagnetic quantum phase transitions is continuous.   Thus this
model fits nicely into the framework discussed in
Sec.~\ref{sec:symm-order-param}:   
For any finite $K/J_1$ up to the critical value, the N\'{e}el and
ferroquadrupolar orders actually coexist, leading to a smooth reduction
in the value of $|\langle{\bf S}\rangle|$ as $K$ is increased.   

\subsection{Separation of Temperature Scales}

\begin{figure}[t]
\includegraphics[width=\columnwidth]{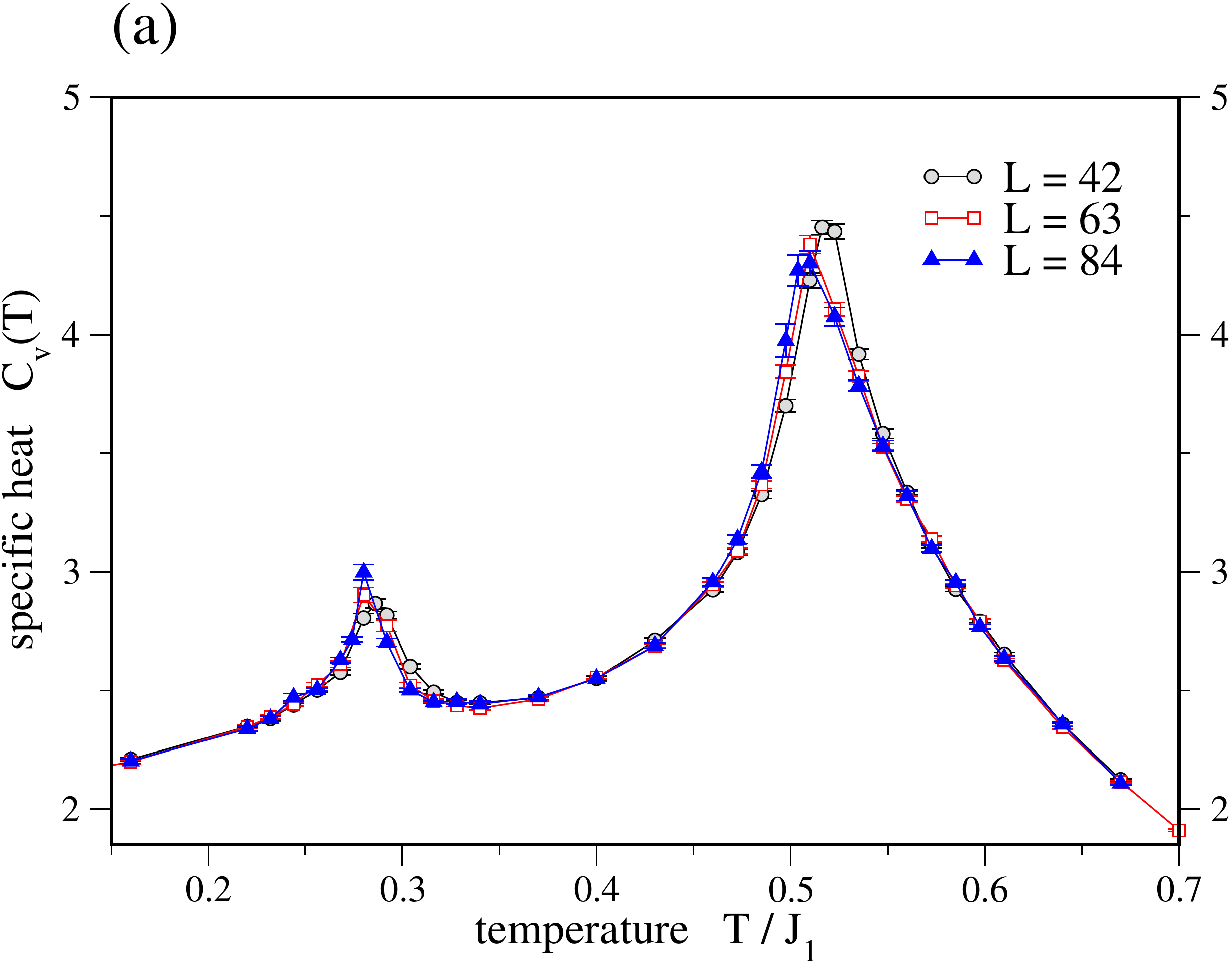}
\includegraphics[width=\columnwidth]{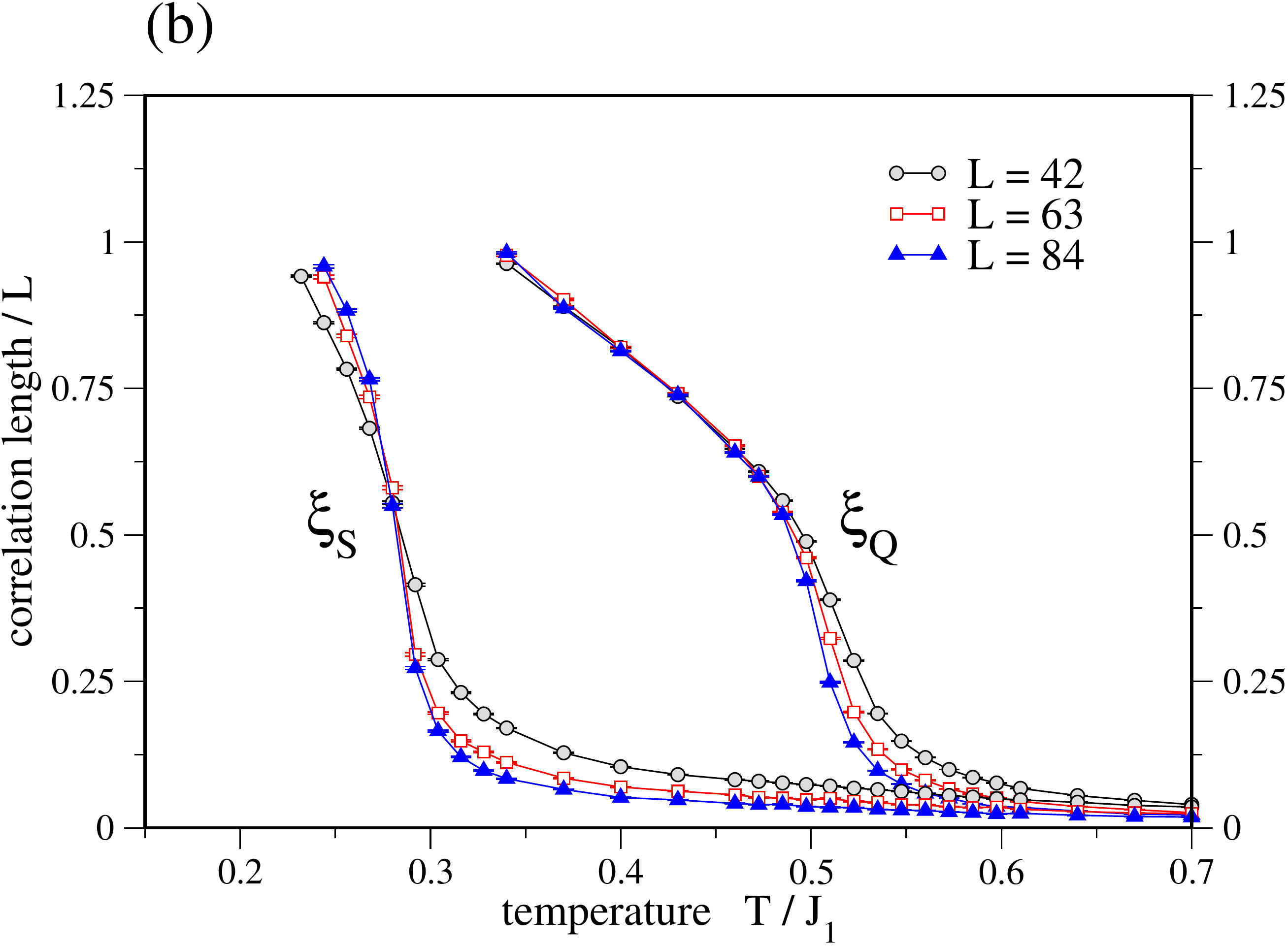}
\caption{
  Finite temperature calculations of (a) the specific heat and 
  (b) the spin and quadrupolar correlation lengths $\xi_S$ and $\xi_Q$
  for the nearest-neighbor model in the semiclassical sSU(3) approximation.
  The two temperature scales associated with the rapid growth of magnetic
  and ferroquadrupolar correlations result in a characteristic two peak
  structure of the specific heat.
  The data shown is for couplings $K/J_1 = 1.5$ and system sizes chosen
  to be commensurate with the low-temperature spin ordering.
}
\label{fig:J1_thermodynamics}
\end{figure}

We have argued above that close to such a quantum
phase transition from spiral to quadrupolar order a spin system may
develop significant quadrupolar correlations at a distinct, higher
temperature scale than the one associated with the rapid growth of
magnetic correlations.  
Our numerical simulations of the sSU(3) model associated with 
Hamiltonian \eqref{eq:j1k-Hamiltonian} does indeed reveal such
a characteristic separation of temperature scales for $K/J_1$
sufficiently close to the quantum phase transition which in mean-field
approximation occurs at  $K/J_1 = 2$. 

In particular, 
we find that in the proximity of the quantum phase transition $K/J_1 < 2$
the measured specific heat develops two distinct peaks
as illustrated in Fig.~\ref{fig:J1_thermodynamics}.
Calculating the ferroquadrupolar and magnetic correlation lengths,
we can associate  the upper peak with the rapid growths of 
ferroquadrupolar correlations while the lower peak indicates the 
onset of magnetic correlations 
(with a wavevector of length $4\pi/3$ that is characteristic of 
the 120-degree state).
We note that although the correlation lengths grow to be very large 
below each of the two temperature scales, they remain finite, 
consistent with the absence of long-range order.
While the high temperature peak in the specific heat shows no
singular behavior as we cross the location of the quantum phase
transition at $K/J_1 = 2$, we find that the lower peak approaches
zero-temperature exactly at the QPT.
For smaller values of $K/J_1 < 2$ the two peaks get closer and
eventually merge around $K/J_1 \approx 5/4$, below which we 
only observe one temperature scale indicated by a single peak
in the specific heat.

We summarize our results for the nearest-neighbor model \eqref{eq:j1k-Hamiltonian}
in the phase diagram of Fig.~\ref{fig:J1_K_two_peak}, where the 
phase boundaries indicate the locations of the respective peaks
in the calculated specific heat.
Note that these phase boundaries are {\sl not} phase transitions, due to the 
absence of long-range order as mandated by the Mermin-Wagner theorem.  

Finally, we note that the numerical results of our sSU(3) approximation
are not expected to be quantitatively correct, as it becomes clear from
the quantitative discrepancy to exact diagonalization results. 
However, we emphasize that the key ingredient underlying our qualitative 
results is a continuous quantum phase transition in which the magnetic order 
smoothly vanishes but the quadrupolar order survives in both phases.
As long as the full inclusion of quantum effects does not lead to the zero
temperature phase transition actually becoming strongly first order, then a
separation of temperature scales should generically be observed when the
parameters of this system are near the quantum critical region.

\section{The $J_3$-$J_1$-$K$ Model}
\label{sec:J3-model}

\begin{figure}[t]
\includegraphics[width=\columnwidth]{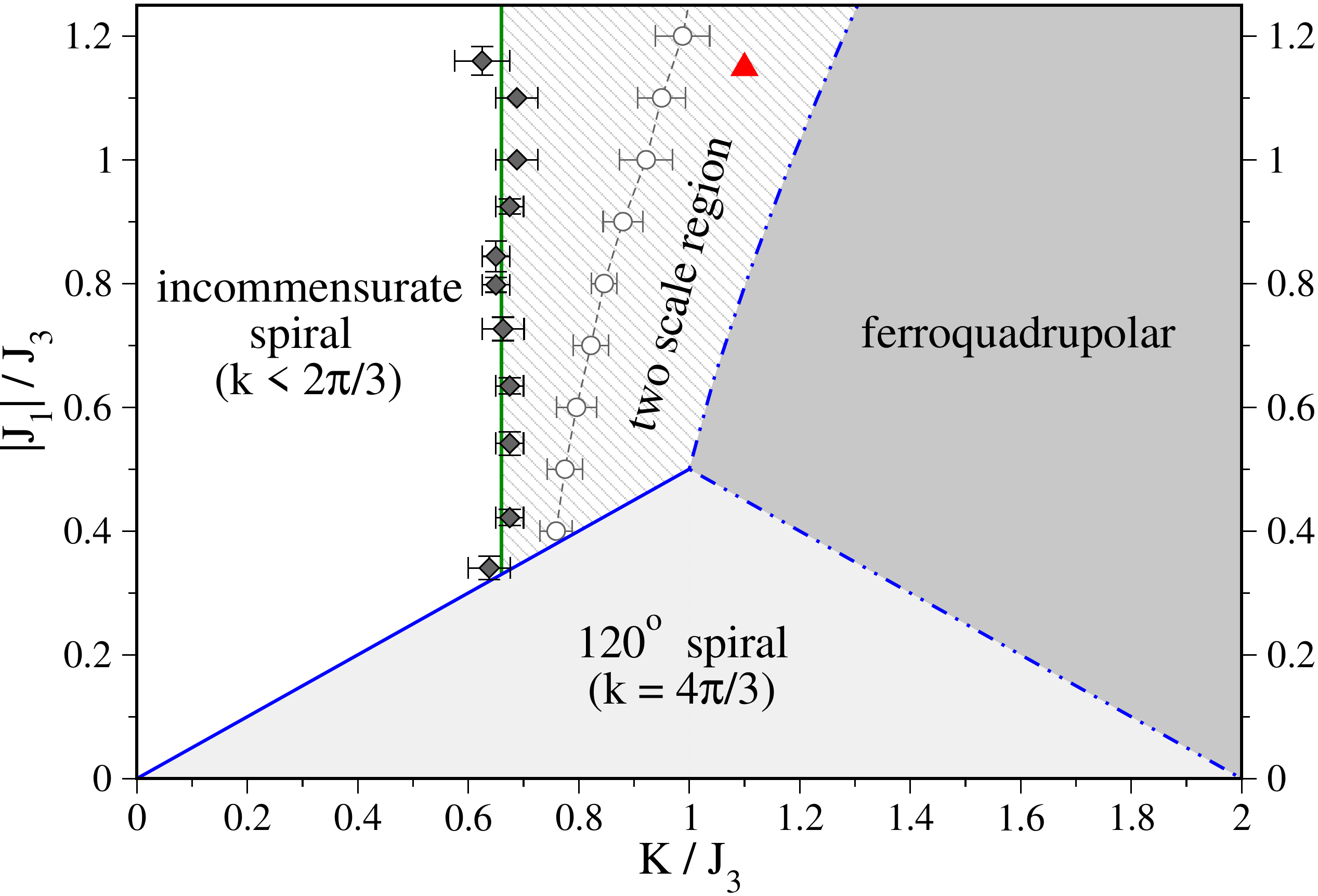}
  \caption{
  Zero temperature phase diagram of the model with antiferromagnetic 
  third-nearest-neighbor coupling $J_3$ and nearest-neighbor ferromagnetic $J_1$ 
  and ferronematic $K$ couplings obtained from numerical simulations in the sSU(3) 
  approximation.
  The three phases present at zero temperature have a ferroquadrupolar, 120 degree spiral 
  or incommensurate spiral ground state, 
  with continuous transitions (dashed lines) between the ferroquadrupolar
  and spiral states and a first-order transition between the 120 degree and incommensurate
  spiral states.  
  The cross-hatched area marks the region where our finite temperature calculations 
  indicate two temperature scales. Such a separation is observed both in specific heat
  (filled diamonds) and correlation length measurements (open circles) which are expected
  to coincide in the thermodynamic limit.
  The red triangle marks the location of the representative system, for which we show
  finite temperature properties in Fig.~\ref{fig:J3_thermodynamics}.
 }
\label{fig:J3_phase_diagram}
\end{figure}

We now turn to the full model in Eq.~\eqref{eq:1} including a third-nearest-neighbor
coupling, which is necessary to properly capture {\sl incommensurate} magnetic
correlations at low temperatures.  We assume antiferromagnetic
third-neighbor interactions $J_3>0$ and ferronematic $K>0$ (as before), 
but now consider a \emph{ferro}magnetic nearest-neighbor coupling $J_1<0$.  
It is this combination of couplings that will realize the proper magnetic correlations
observed in \NiGaS.
As we shall see, the presence of an incommensurate spiral ground state has interesting 
consequences in that it introduces a true finite temperature phase transition associated
with anisotropic spin fluctuations.

While we find that the above model can now account for the type of magnetic
order observed in \NiGaS, it still cannot explain the fact that the spin
order never grows to be long ranged and exhibits slow
dynamics. Therefore we will ultimately have to consider the role of
disorder before discussing our results in the context of the real material.

\subsection{Ground States}

We again start our analysis of of Hamiltonian \eqref{eq:1} by discussing
its zero temperature phase diagram calculated in mean-field approximation.
For the range of parameters of interest, we assume that it is sufficient to take 
the magnetic ground states to be coplanar and invariant under the combination 
of a lattice translation and spin rotation associated with a single wavevector.
Under these two assumptions, the most general possible choice for the
vector of amplitudes ${\bf b}_i$ describing the wavefunction at a single
site is, up to a global O(3) spin rotation,
\[
{\bf b}_i = \left[
\begin{array}{l}
\sin\theta \cos({\bf k}\cdot{\bf r}_i) e^{i\alpha}\\
\sin\theta \sin({\bf k}\cdot{\bf r}_i) e^{i\alpha}\\
\cos\theta
\end{array}
\right] \,.
\]
Minimizing the energy over the variables ${\bf k}$, $\theta$ and $\alpha$ we obtain
the zero temperature phase diagram shown in Fig.~\ref{fig:J3_phase_diagram}. 
For small values of $K/J_3$, the ground states are again spirals with a reduced spin 
moment. 
Along the line $|J_1| = K/2$ for $K<1$, the wavenumber describing these spiral states 
ground states jumps from $4\pi/3$ to $2\pi/3$ and then continuously decreases to zero 
with increasing $|J_1|$, ultimately leading to a ferromagnetic ground state for very large 
values of $|J_1|/J_3$.
The new feature in this phase diagram is thus a large region of parameters for which the 
spiral is incommensurate (with wavevector $k < 2\pi/3$).
Increasing the biquadratic exchange $K$ for fixed $|J_1|/J_3$, these spiral ground states 
smoothly transform into a ferroquadrupolar ground state with $\avg{{\bf S}} = 0$ which is
the ground state for large $K$. 

\subsection{Separation of Temperature Scales}

Similar to the nearest-neighbor only model we expect on symmetry grounds that also the
model with third-neighbor interactions exhibits a separation of temperature scales in the
vicinity of the continuous quantum phase transition between the ferroquadrupolar and 
(incommensurate) spin spiral state.
Indeed our numerical simulations of this model in the sSU(3) approximation reveal a
broad range of couplings in the vicinity of this line of continuous transtions 
where both specific heat and correlation
length measurements indicate two different temperature scales as indicated by the 
cross-hatched area in Fig.~\ref{fig:J3_phase_diagram}.

\begin{figure}[htp]
\includegraphics[width=\columnwidth]{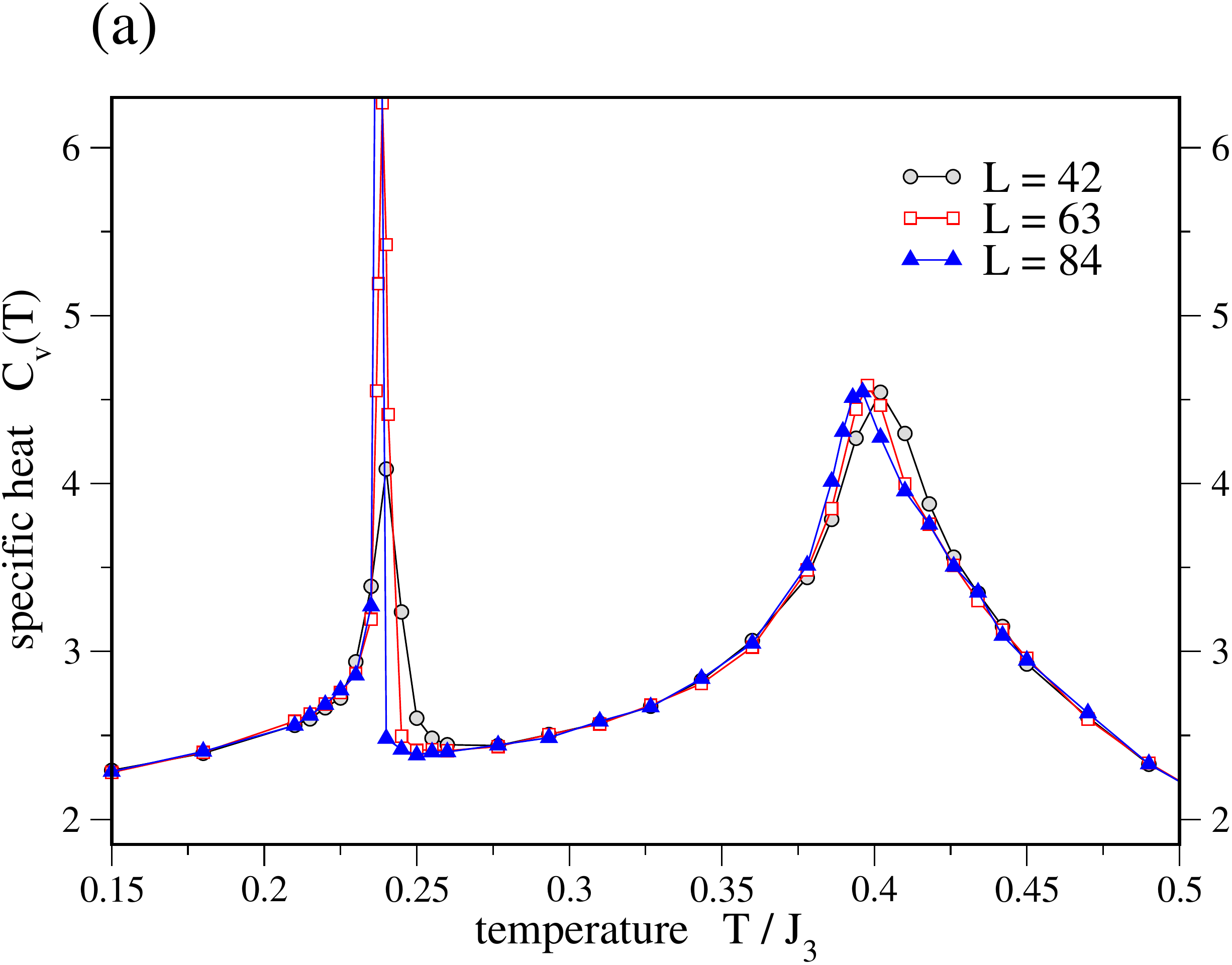}
\includegraphics[width=\columnwidth]{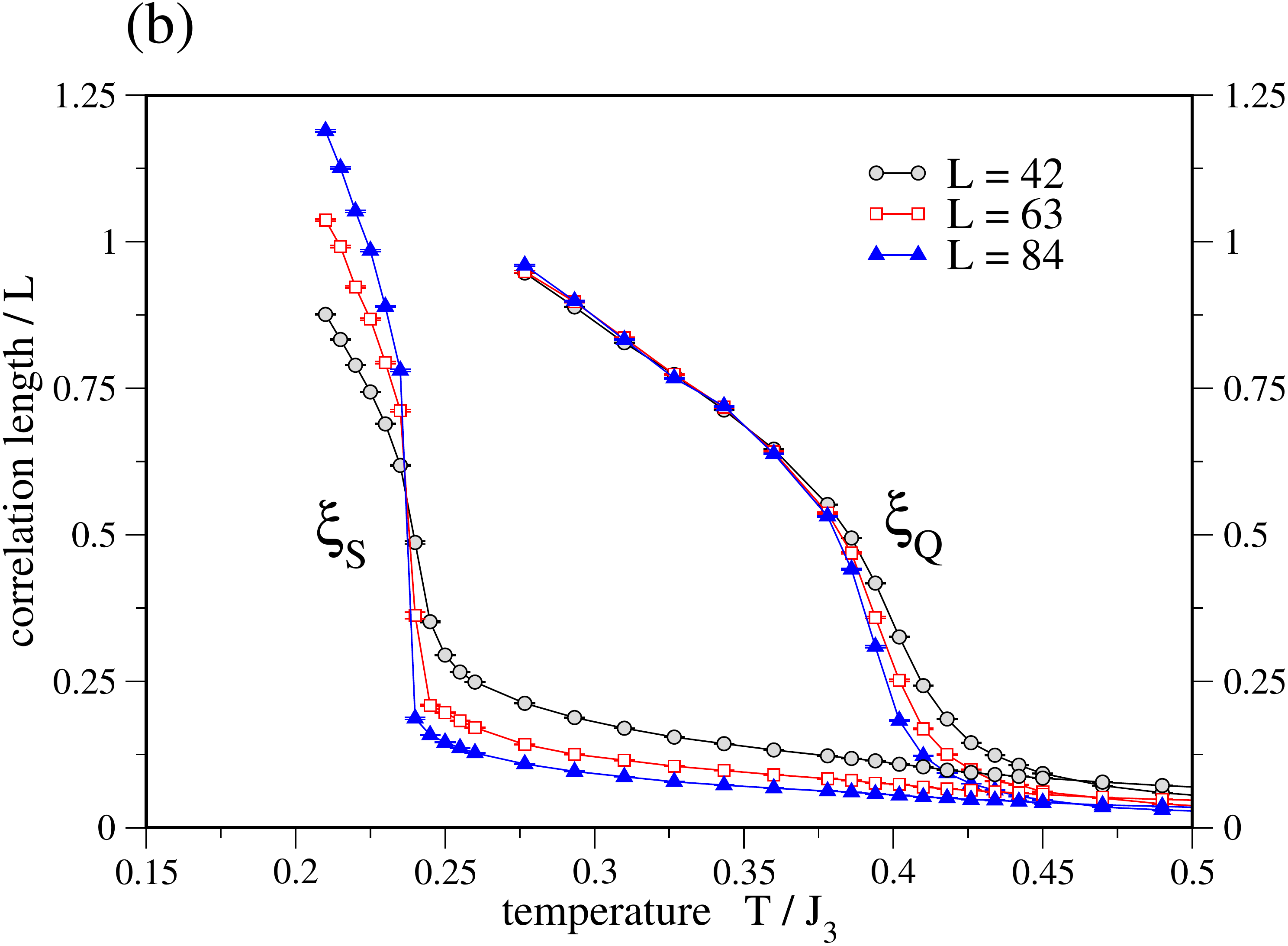}
\includegraphics[width=\columnwidth]{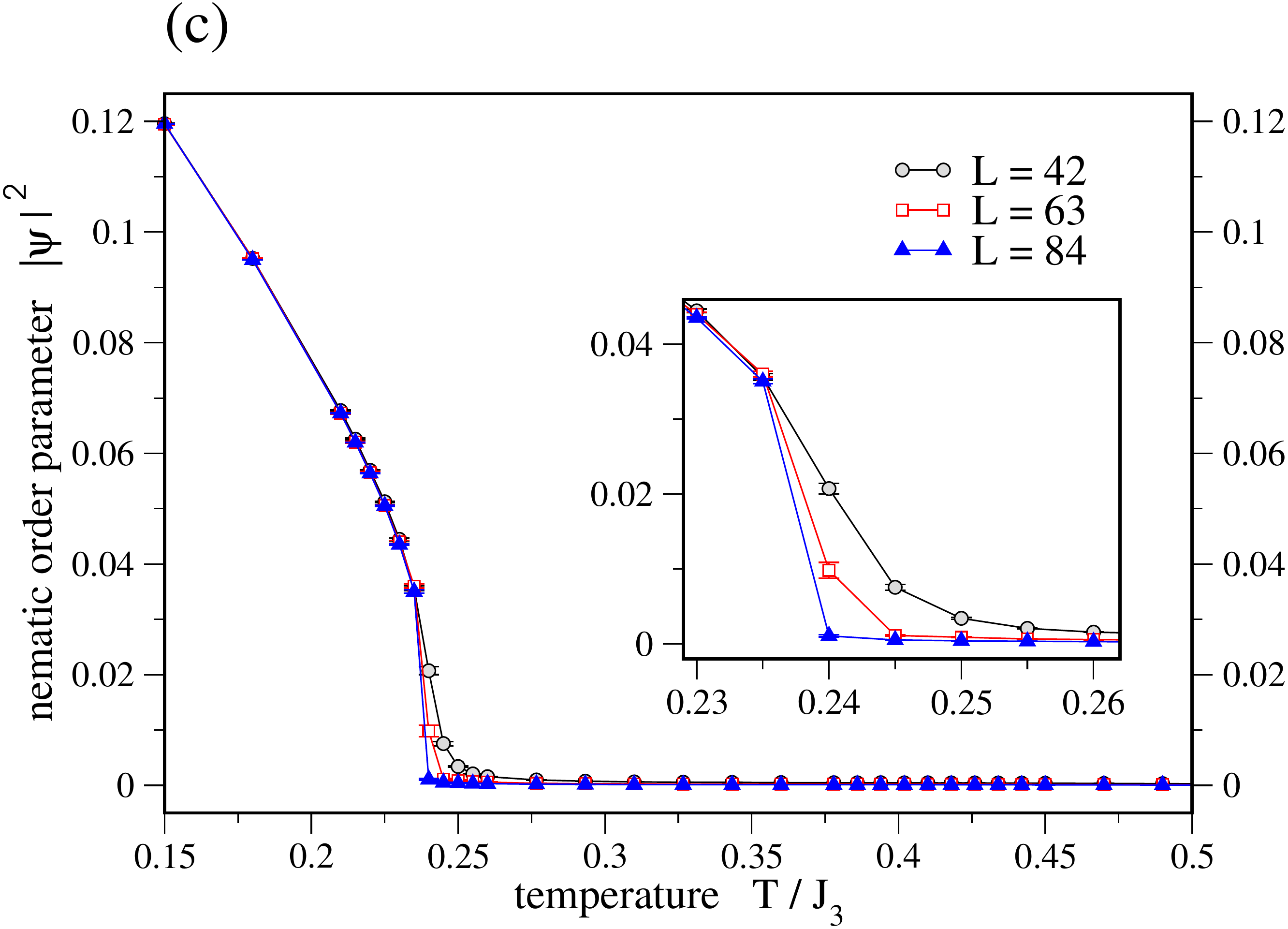}
\caption{
  Finite temperature calculations of (a) the specific heat, 
  (b) the spin and quadrupolar correlation lengths $\xi_S$ and $\xi_Q$ and
  (c) the nematic order parameter $|\psi|^2$
  for the model with third-nearest-neghbor interaction in the semiclassical 
  sSU(3) approximation.
  The two temperature scales associated with the rapid growth of magnetic
  and ferroquadrupolar correlations again result in a characteristic two peak
  structure of the specific heat.
  The data shown is for couplings $K/J_3 = 1.1$ and $J_1/J_3 = -1.14855$
  and system sizes chosen to be commensurate with the low-temperature spin ordering.
  In contrast to the nearest-neighbor model, we see a divergent lower peak corresponding to the 
  breaking of lattice rotational symmetry by anisotropic spin fluctuations. 
}
\label{fig:J3_thermodynamics}
\end{figure}

In agreement with the symmetry arguments in support of this temperature scale separation
we again find that the upper temperature scale corresponds to the rapid growth of quadrupolar correlations while the lower temperature is associated with the onset of magnetic correlations. 
Numerical data for a representative choice of parameters (indicated by the red triangle in
Fig.~\ref{fig:J3_phase_diagram}) is shown in Fig.~\ref{fig:J3_thermodynamics}, where we plot
the specific heat and correlation length measurements as before. The exact parameters
$K/J_3 = 1.1$ and $J_1/J_3 \approx -1.14855$ and system sizes are again chosen such that
the low temperature spiral state has a commensurate spatial period (in this case 7 lattice spacings) which in turn allows to compute the correlation lengths directly from the respective structure factors.

\begin{figure}[t]
\includegraphics[width=\columnwidth]{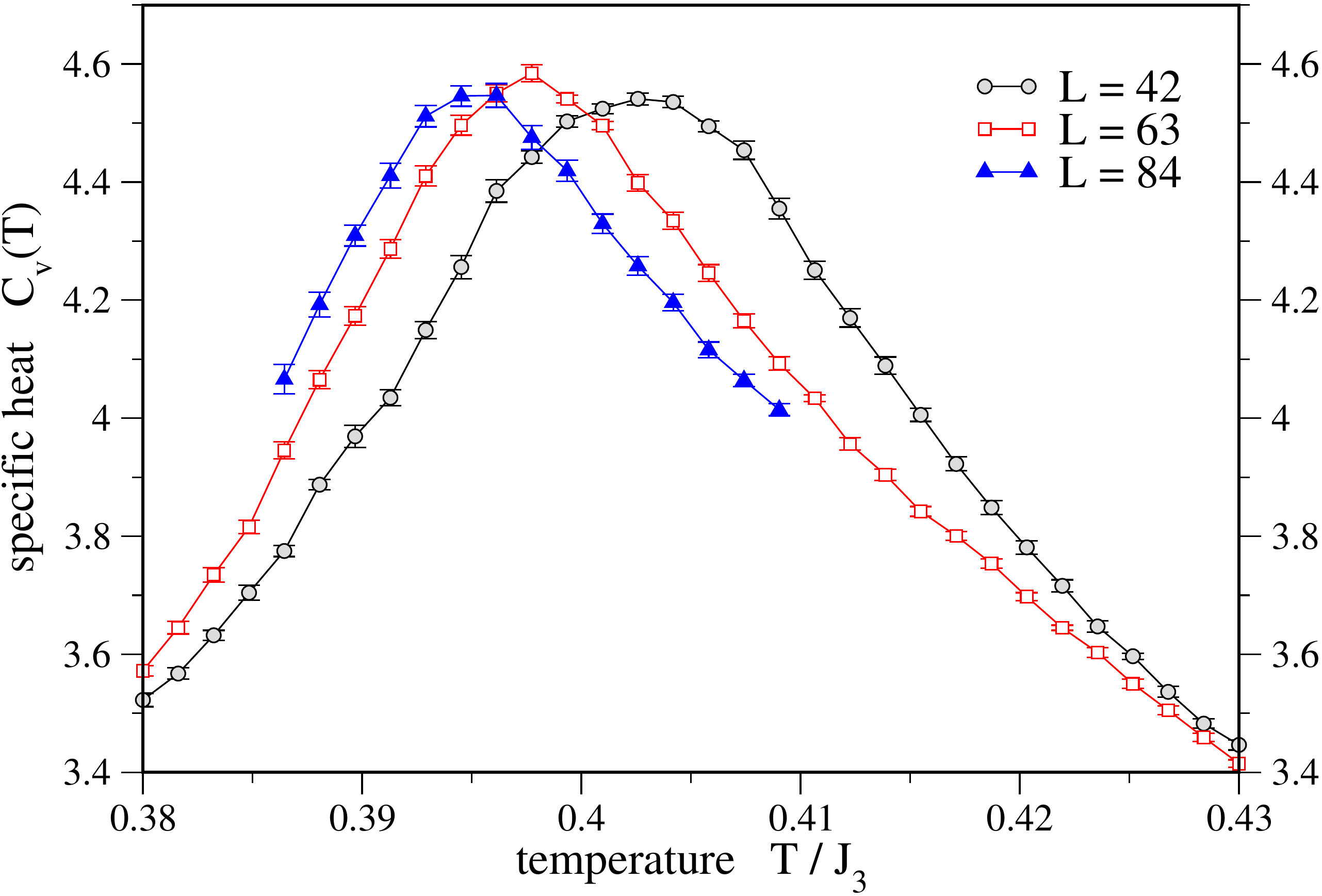}
\caption{
  Finite-size scaling of the upper peak in the specific heat of the representative system 
  with couplings $K/J_3 = 1.1$ and $J_1/J_3 = -1.14855$. 
  While the peak slightly shifts with increasing system size, there is clearly no divergence.
}
\label{fig:upk_scaling}
\end{figure}

\begin{figure}[t]
\includegraphics[width=\columnwidth]{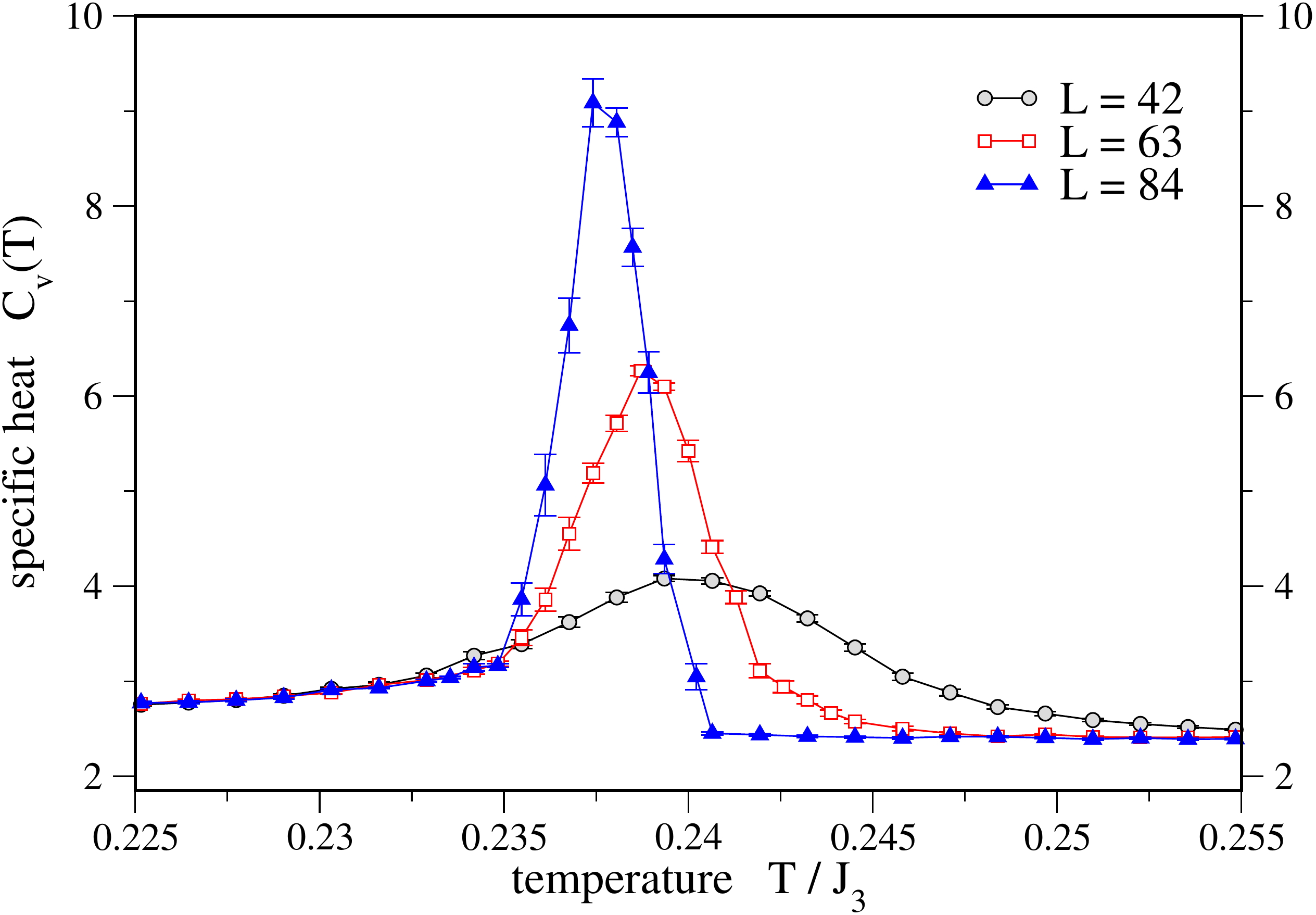}
\caption{
  Finite-size scaling of the lower peak in the specific heat of the representative system 
  with $K/J_3 = 1.1$ and $J_1/J_3 \approx -1.14855$. The trend with increasing system
  size clearly suggests a divergence in the thermodynamic limit.
}
\label{fig:lpk_scaling}
\end{figure}

\subsection{C$_3$ Bond Ordering Transition}

\begin{figure}[t]
\includegraphics[width=\columnwidth]{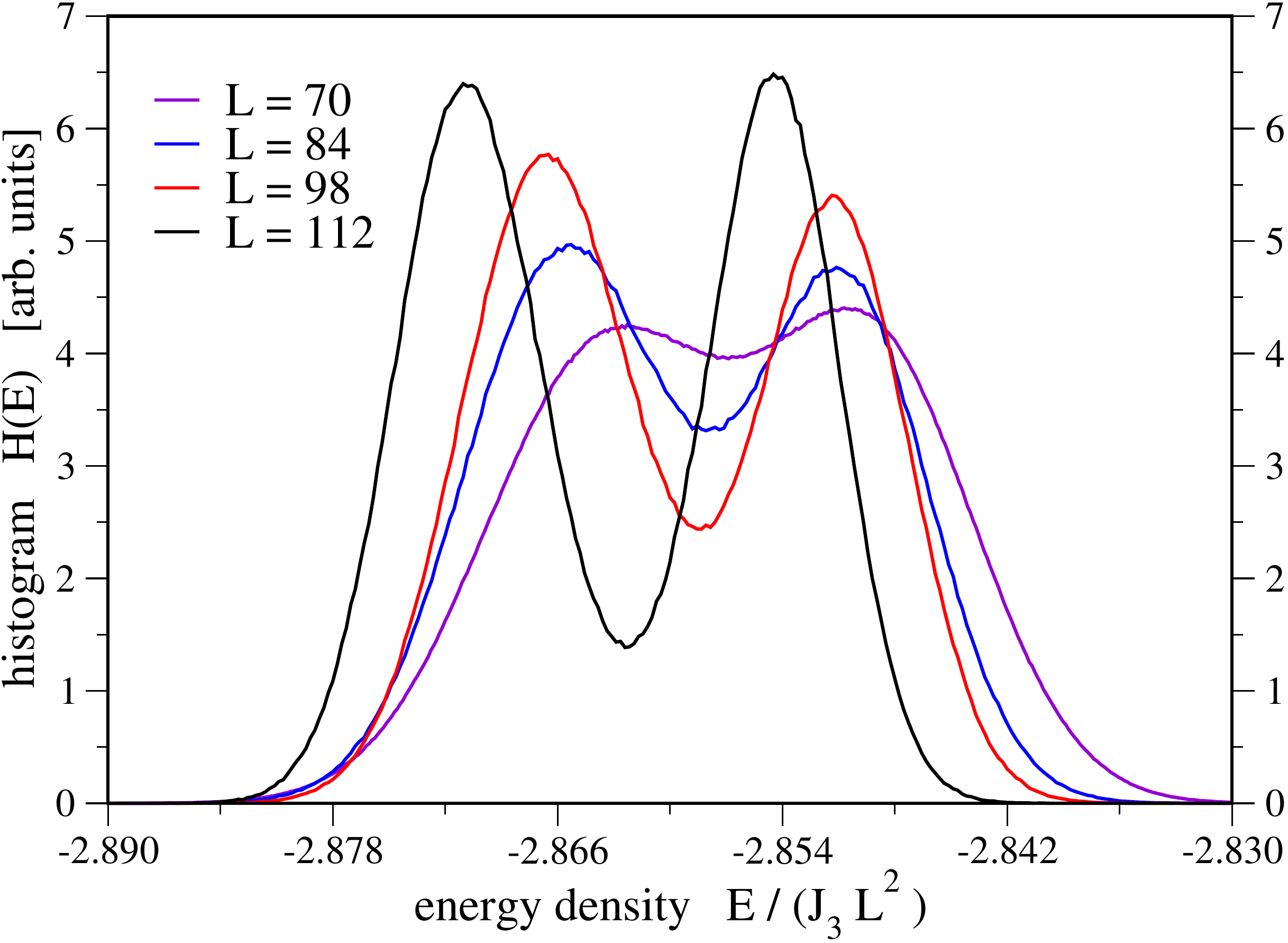}
\caption{
  Histogram of the energy density for our representative system in the vicinity
  of the low temperature transition.
  The bimodal character of this distribution proliferates with increasing system
  size, which is indicative of a first-order transition.
  Data for different system sizes/temperatures are shown;  in particular
  $L = 84$ ($T/J_3 = 0.2377$), 
  $L = 98$ ($T/J_3 = 0.2374$), and $L = 112$ ($T/J_3 = 0.2361$).
}
\label{fig:histogram}
\end{figure}

As ferroquadrupolar correlations only break the continuous SU(2) symmetry, 
the transition at the upper temperature scale should not be associated with the 
onset of true long-range order as mandated by the Mermin-Wagner theorem.
Indeed a finite-size scaling analysis of our numerical data shows that the
high-temperature peak in the specific heat is non-divergent as we increase
the system size, as illustrated in Fig.~\ref{fig:upk_scaling}.

A similar analysis for the transition at the lower temperature scale, however, 
reveals that more interesting things can happen. For any choice of parameters 
that puts the ground state of the system into the incommensurate spiral phase, 
the lower peak of the specific heat appears to diverge sharply as shown in
Figs.~\ref{fig:J3_thermodynamics}(a) and \ref{fig:lpk_scaling}. 
This divergence indicates a true phase transition, which in two dimensions must 
reflect the breaking of some discrete symmetry in accordance with the 
Mermin-Wagner theorem. (In principle such a divergence could also be
explained by a first-order transition without a change of symmetry, but will be 
ruled out below.)

The key observation to explain this phase transition to a long range ordered
state is that any (incommensurate) spiral state on the triangular lattice with
wavenumber $|k| < 4\pi/3$ exhibits spin correlations that are stronger along 
one axis than the other two, unlike the 120-degree state, in which every spin 
differs from its neighbors by the same angle.
Therefore any incommensurate spiral state spontaneously breaks the discrete 
$C_3$ rotational symmetry of the lattice and there is nothing to prevent
this symmetry  breaking from occurring at a finite temperature.  
Indeed, it has recently been shown that the related classical model with O(3) spins 
on the triangular lattice and first and third neighbor exchange spontaneously
breaks the $C_3$ lattice rotation symmetry at low temperature~\cite{JPSJ.77.103002}.
In the $C_3$ broken state the system exhibits ``bond order'' in the sense that the 
exchange energies for bonds along different principle axes become distinct.  

To confirm that the lower temperature scale in our sSU(3) model does indeed 
correspond to the breaking of lattice rotational symmetry, we define a bond order 
parameter $\psi$, which is nonzero only in the presence of anisotropic spin correlations:
\[
\psi = \frac{1}{N} \sum_{i,\mu} (a^x_\mu + i a^y_\mu ) \, \avg{{\bf
    S}_{{\bf r}_i}\cdot{\bf S}_{{\bf r}_i + {\bf a}_\mu}} \,,
\]
where ${\bf a}_1 = \hat{x}$ and ${\bf a}_{2,3}$ are its rotations by
$2\pi/3$ and $4\pi/3$, respectively.  This order parameter is a complex
number whose argument reveals the direction of the strongest
correlations and whose magnitude for a spiral state with wavenumber $k$
is given by
\[
|\psi| = \avg{|{\bf S}|}^2\, [\cos(k)-\cos(k/2)]\ .
\]
Moreover, we note that $\psi$ is invariant under the transformation
${\bf S}_i \rightarrow -{\bf S}_i$ and under a lattice rotation of angle
$\pi$.  Thus, any nonzero value of $\psi$ indicates that the system is
one of three inequivalent bond ordered phases related to each
other by lattice rotations of $2\pi/3$ (cf. the discussion on spiral
magnetic domains in section \ref{sec:symm-order-param}).

Measurements of this bond order parameter for the sSU(3) model associated with 
Hamiltonian \eqref{eq:1} confirm that the system enters a bond ordered state for 
couplings leading to an incommensurate spiral ground state, as shown in 
Fig.~\ref{fig:J3_thermodynamics} (c) for our representative system.
To determine the order of this low temperature transition we have measured
energy histograms in the vicinity of the transition temperature. The bimodal
histogram distribution, shown in Fig.~\ref{fig:histogram}, unambiguously points
to a first-order transition, which is also indicated by the narrow peak in the
specific heat and the apparent jump of the magnetic correlation length in 
Fig.~\ref{fig:J3_thermodynamics} (b).
Thus the C$_3$ transition in the sSU(3) model has the same qualitative
properties as the corresponding transition in the O(3) model which was also
found to be first order \cite{JPSJ.77.103002}.  

\section{Disorder}

\subsection{Imry-Ma argument and correlation lengths}
\label{sec:imry-ma-argument}

We now turn to the effects of disorder.  We consider sulfur (S)
vacancies to likely be the dominant type of impurity in \NiGaS.
However, our analysis will be rather general, and rely only upon the
assumption that the defects are uncorrelated, local, and
non-magnetic. On symmetry grounds, non-magnetic disorder couples
directly to the $C_3$ order parameter, but not to ${\bf d}$, since
(neglecting spin-orbit) it respects SU(2) symmetry.  Thus for the $C_3$
order, the impurities act as ``random fields''.  In two dimensions, the
standard Imry-Ma argument implies that long-range $C_3$ order is
destroyed by arbitrarily weak
disorder.\cite{nattermann97:_theor_of_random_field_ising_model}\ The
system breaks up into ``domains'', leading to a saturation of
 the correlation length $\xi_{C_3}$ for the $C_3$ bond order.  We
parametrize the strength of the random field by an energy $\Delta$,
which gives e.g. the range of variation in strength of exchange
couplings due to S vacancies. The essence of the Imry-Ma argument is a
competition between the surface energy required to nucleate a domain of
one phase within another, and the energy gain that can be obtained from
the random fields within the domain.  The former is proportional to the
surface tension $\sigma$ of a domain, while the latter is proportional
to $\Delta |\psi|$.  Hence we expect $\xi_{C_3}$ to be a function of
$\sigma/(\Delta |\psi|)$.  According to the Imry-Ma
argument,\cite{nattermann97:_theor_of_random_field_ising_model}\ two
dimensions is the marginal (lower critical) dimension for the stability
of the ordered state.  Hence the correlation length is an {\sl
  exponential} function of the disorder strength.  We arrive at the
conclusion
\begin{equation}
  \label{eq:8}
  \xi_{C_3}(T) \sim \exp\left[ \frac{ c \sigma(T)^2}{\Delta^2
      |\psi(T)|^2}\right],
\end{equation}
with some constant $c$.  At $T\ll T_{C_3}$, we expect
$\sigma(0) \sim O(J_3)$ and $\psi(0) \sim O(1)$. Eq.~(\ref{eq:8})
describes a continuous decrease of the $C_3$ correlation length on
increasing temperature up to $T\approx T_{C_3}$, the temperature of the $C_3$
transition of the pure system.  At this point, the first order
transition of the pure system will be rounded by disorder, and the $C_3$
correlation length will smoothly (and quickly) approach its finite value
for the pure system for $T>T_{C_3}$.  

What of the magnetism?  Although the magnetic order parameter does not
couple direction to disorder, it will be strongly affected by the domain
structure.  In particular, the conventional spiral correlation length
will saturate at the domain size,
\begin{equation}
  \label{eq:6}
  \xi_M(\Delta,T) \approx \min\, \left[ \xi_M(0,T),
    \xi_{C_3}(\Delta,T)\right]. 
\end{equation}
Because the pure system's magnetic correlation length grows very large
below $T_{C_3}$, we expect that for most if not all of this temperature
range, that simply $\xi_M=\xi_{C_3}\equiv \xi$.

\subsection{Correlations on longer scales: gauge glass physics}
\label{sec:corr-long-scal}

The finiteness of the above correlation lengths does not imply a complete
absence of magnetic correlations beyond this length scale, only that the
{\sl spiral} correlations remain short range.  One longer scales, we
need to consider more subtle correlations between spins in different
domains.

In what follows, we will make a simplifying assumption, that the spirals
are fixed in the XY plane.  This could be justified by a small
easy-plane anisotropy, $H_{\rm anis} = D \sum_i \left( S_i^z\right)^2$,
which is expected to be present, but small, in \NiGaS.  Provided $D
\xi^2 \gtrsim k_B T$, which will be satisfied at low enough
temperatures, the spins will indeed lie in the XY plane.  It may also be
the case that the spins adopt a coplanar state spontaneously, even in
the absence of magnetic anisotropy.  However, we do not make a
definitive statement in this respect.  Should this assumption fail, we
expect that this approximation {\sl underestimates} the fluctuations,
and therefore overestimates the tendency to spin freezing.  One should
probably therefore view the estimates of the spin glass correlation
length and time below as upper bounds.

To proceed, we define ${\bf d}_i=({\bf\hat x}+i {\bf\hat
  y})e^{i\theta_i}$ as the average order parameter for a domain $i$,
within which the wavevector is fixed at ${\bf Q}_{b_i}$.  In general,
the energy of the system will depend upon the difference of the angles
$\theta_i-\theta_j$ in adjacent domains, due to the coupling of spins at
the domain boundaries.  This coupling is complicated and depends upon
the specific structure of this boundary, and the wavevectors of the two
domains.  Therefore we will assume the effective Hamiltonian
\begin{equation}
  \label{eq:7}
  H_{\rm eff} = -K \sum_{\langle ij\rangle}  \cos(\theta_i-\theta_j -
  \alpha_{ij}), 
\end{equation}
where $\alpha_{ij}$ is a random variable, which we take to be
independent on each $ij$ and uniformly distributed between $0$ and
$2\pi$.  We have neglected for simplicity randomness in the magnitude of
the coupling between domains, which does not qualitatively affect the
physics.  The coefficient $K$ arises from adding the exchange couplings
along the boundary of the two domains in question.  If the addition is
constructive, we expect, ignoring fluctuations $K\sim J_3 \xi$, or if it
is random instead $K \sim J_3 \sqrt{\xi}$.  With thermal and/or quantum
fluctuations, this should be reduced by a factor of $\langle
|d|^2\rangle$.  Assuming a mean-field temperature dependence of the
magnitude of the local spiral order parameter, one finds a smooth
reduction with increasing temperature $K(T) = K(0) (1- T/T_{MF})$, for
$T<T_{MF}$.  Generally, $K \gg k_B T$ in the regime where we can apply
Eq.~(\ref{eq:7}).

Eq.~(\ref{eq:7}) is the Hamiltonian of the two-dimensional {\sl gauge
  glass} model,  which has been intensively studied in the past.\cite{PhysRevB.38.386,PhysRevB.66.224507,PhysRevB.69.184512,PhysRevB.43.130}  
It is believed that its lower critical dimension is greater
than two, so that the system is disordered (paramagnetic) at all
$T>0$.\cite{PhysRevB.69.184512}  However, the correlation length for
glassy (Edwards-Anderson) order, and the correlation times of the spins, diverge
rapidly as $T\rightarrow 0$.  

To understand this behavior in more detail, we must discuss the
excitations of the gauge glass ground state.  Obviously, because of the
global $U(1)$ symmetry of Eq.~(\ref{eq:7}), there are Goldstone-type
spin waves, in which the phases $\theta_i$ vary slowly away from their
ground state values.  These are the Halperin-Saslow modes, introduced in
Ref.~\onlinecite{halperin1977hts} and discussed in 
Ref.~\onlinecite{podolsky2008hsm} in the context of \NiGaS.  More subtle,
however, are non-smooth deformations of the phase, which have {\sl
  lower} energy than the spin waves for a corresponding length scale.
These are the ``droplets'' of the droplet theory.\cite{PhysRevB.38.386}  From numerical
studies, these have a classical energy $E \sim A L^{-|\theta|}$, with
$\theta \approx - 0.36$.\cite{PhysRevB.69.184512,PhysRevB.66.224507}  Here $L$ should be measured in units of the
effective lattice spacing of Eq.~(\ref{eq:7}), and $A$ should be of
order $K$.  Thus at low temperature, those ``droplets'' with size larger
than the glass correlation length,
\begin{equation}
  \label{eq:9}
 \xi_g \sim \xi (T_0/T)^{1/|\theta|},
\end{equation}
will be thermally activated, and presumably fluctuating in and out.
Here $k_B T_0 \sim K$.  
These droplets should be responsible for the vanishing of the static
moment.  

The question is, what is the time scale for these fluctuations?  This is
generally determined by the energy barrier which must be overcome to
create one of the droplet excitations (these can also be thought of
in terms of moving vortices in some region of size $\xi_g$).  According
to scaling theory, this is of order $B \xi_g^\psi$, with some ``barrier
exponent'' $\psi$.  This gives the general expectation
\begin{equation}
  \label{eq:11}
  \tau_g \sim \exp \left[ (T_0/ T)^{1+\psi/|\theta|}\right].
\end{equation}
Some recent numerics\cite{PhysRevB.69.184512}
suggest $\psi=0$, so that the barriers are logarithmic, of order $B \ln
\xi_g$ (again $B \sim K$).  In this case, the glass correlation time
behaves like
\begin{equation}
  \label{eq:10}
  \tau_g \sim \exp [ B \ln \xi_g /k_B T] \sim \exp\left[ \frac{T_0 \ln
      (T_0/T)}{T}\right].
\end{equation}
Both Eq.~\eqref{eq:11} and Eq.~\eqref{eq:10} describe super-Arrhenius
divergence of the relaxation time at low temperatures. 
For times shorter than this correlation
time, the spins are effectively frozen.  Due to the rapid growth of the
correlation time at low temperature, this gradual process can be easily
misinterpreted as a sharp dynamical transition.  Nevertheless the actual
freezing within the gauge glass model is a gradual and smooth process.
Specifically, dependent upon the probe used, we expect that persistent
spin dynamics can be observed below a nominal  ``freezing'' temperature
(most likely the latter, empirically determined, temperature will be close
to $T_{C_3}$).

\section{Discussion}
\label{sec:discussion}

\subsection{Summary}
\label{sec:summary}

In this paper, we have studied an extended bilinear-biquadratic exchange
model for an $S=1$ antiferromagnet on the triangular lattice, focusing
on $T>0$ properties.  We found that, when the system is proximate to a
$T=0$ quantum phase transition to a ferro-quadrupolar (spin-nematic)
state, the system is characterized by two energy scales.  The specific
heat then displays separate peaks at the two corresponding temperatures.
We introduced a novel sSU(3) approximation and associated numerical
method to study such $T>0$ properties.  Below the lower temperature
peak, long-range $C_3$ bond order is present, and spiral magnetic
correlations rapidly develop.  In this range of temperature, the system
is exquisitely sensitive to disorder, which acts like a ``random field''
on the $C_3$ order parameter.  We argued that this results in a gradual
freezing phenomena, related to the well-studied ``gauge glass'' model,
with persistent but rapidly slowing dynamics down to zero temperature.

\subsection{Relation to experiments}
\label{sec:relation-experiments}

\subsubsection{Model}
\label{sec:model}

Having established these theoretical results, we now discuss their
relevance to \NiGaS.  First, we consider the appropriateness of the
model Hamiltonian, Eq.~(\ref{eq:1}).  Neutron experiments have
decisively established the nature of short-range spin correlations.
Their location near the $(\tfrac{1}{6},\tfrac{1}{6},0)$ point in the
Brillouin zone clearly implies the dominance of third neighbor
antiferromagnetic exchange $J_3$ over first and second neighbor
exchange.  Classically, weak ferromagnetic $J_1$ selects this ordering
wavevector and accounts for the slightly incommensurate value.  This is
also true quantum mechanically: for a
model involving only $J_1$ and $J_3$, with $J_1 \ll J_3$, it is possible
to confirm this type of spiral ground state directly for $S=1$, assuming
only that a nearest-neighbor triangular antiferromagnet with $S=1$ is
ordered.\cite{m.:_unpub}\ 
This pattern of exchange interactions has also been justified by microscopic
considerations.\cite{mazin-07,takubo:037203}\ On physical grounds,
this situation implies an anomalously low overlap in the
nearest-neighbor Ni-S-Ni superexchange pathway, which involves a
90$^\circ$ bond.  According to the usual Goodenough-Kanamori rules, such
a pathway indeed results in a weak ferromagnetic exchange.  However,
this situation is very sensitive to variations in the bond angle.  We
may therefore expect virtual fluctuations of phonons which deform
this bond angle to make an important contribution to the effective
magnetic Hamiltonian.  In the limit of a high energy phonon (with $\hbar
\omega_{\rm phonon} \gtrsim J_1$), this directly leads to the
effective ferro-biquadratic coupling $K$ in Eq.~(\ref{eq:1}).   

\subsubsection{Intermediate energy physics}
\label{sec:interm-energy-phys}

Is there any empirical evidence for interactions, such as $K$, beyond
Heisenberg exchange?  Though much of the emphasis, both theoretically
and experimentally, on \NiGaS\  has been on its low temperature (below say
10K) behavior, the high temperature phenomena (say for 10K $<T<$ 100K) is
perhaps more anomalous.  In particular, the two peak structure of the
magnetic specific heat, with no indications of a sharp phase transition
at {\sl either} temperature, is quite unusual.  Moreover, the very high
temperature of the upper peak, comparable to the Curie-Weiss temperature
itself, is especially striking for a frustrated magnet, in which the
entropy usually remains down to temperatures well below $\Theta_{CW}$.
Moreover, Ga nuclear magnetic resonance (NMR) and muon spin resonance
($\mu$SR) experiments also show strong temperature dependence of the
dynamics in this broad temperature range.  In our opinion, these
observations strongly suggest that a simple picture in which only
``classical'' frustration is operative is inadequate.  Indeed, it is
natural, given the coincidence of the Curie-Weiss temperature and the
upper specific heat peak, to think that some {\sl unfrustrated} type of
correlations are at work.  The ferro-quadrupolar order parameter
discussed here is our proposed candidate, and we take the two-peak
structure of the specific heat as an indication of the proximity of the
ground state to a $T=0$ quantum phase transition to a non-magnetic
quadrupolar phase.  This is fully consistent with the presence of
non-negligible $K>0$.

It would be highly desirable to elaborate more quantitatively upon this
scenario.  It is tempting to apply directly the Monte Carlo results of
this paper within the sSU(3) approximation to fit data on \NiGaS.
However, we believe this is unjustified, because the sSU(3)
approximation is known to be {\sl quantitatively} bad.  As shown in
Ref.~\onlinecite{laeuchli-06}, at $T=0$ it drastically overestimates the biquadratic
exchange $K$ required to stabilize the ferro-quadrupolar phase in the
nearest-neighbor model.  Significantly more theoretical work is required
to improve on the sSU(3) approximation for the extended exchange model
appropriate for \NiGaS.  The low frequency dissipative susceptibility,
$\chi''(q,\omega)$, relevant to the NMR and $\mu$SR experiments, is
particularly worthy of study.

\subsubsection{Low temperature freezing}
\label{sec:low-temp-freez}

Experiments have uncovered a complex freezing process in \NiGaS.  The
temperature $T_f\approx 10K$ is identified as a ``freezing temperature''
at which there is an apparent divergence of the $1/T_1$ and $1/T_2$
relaxation rates in NMR and NQR measurements.\cite{yaouanc:092403,citeulike:2437587,takeya2008sda}  However, static moments
are {\sl not} formed at this temperature, and several measurements
indicate that spin dynamics persists down to temperatures of at least
$2K$.  For $T<2K$, broad NQR and NMR spectra are observed, as is a 1/3
``tail'' in the $\mu$SR spectrum; both are indicators of inhomogeneous
moments that are static on times of at least the (slowest) muon lifetime
of $\tau_\mu \approx 2.2\mu$sec.  Relaxation of the 1/3 tail is observed for $T>2K$,
indicating that here the spin dynamics is faster than $\tau_\mu$.  This
is consistent with the lack of an NQR signal for $2K<T<T_f$, which
indicates that the spin dynamics has substantial weight at the NQR
frequency.  

In our model, the low temperature specific heat peak is associated with
the onset of substantial magnetic correlations and $C_3$ bond order.
Experimentally, this coincides approximately with $T_f\approx 10K$.
Around this temperature, we expect an increase of the magnetic
correlations towards their low temperature saturation value determined
by the Imry-Ma disorder physics.  As temperature is lowered below this
value, we cross into the gauge glass regime.  Therefore one expects in
this range a gradual freezing process with spin dynamics governed by
Eqs.~(\ref{eq:11},\ref{eq:10}).  This growth of glassy relaxation time $\tau_g$
appears consistent with the experimental observations.  In this picture,
there is no true transition at the experimental $T_f$.  Rather, what is
observed experimentally are the residual effects of the true transition
that would occur around this temperature in an ideal clean sample.

\subsubsection{Why \NiGaS ?}
\label{sec:why-nigas}

A number of aspects of the experimental behavior of \NiGaS\ appear
rather unique among frustrated magnets, including a variety of other
triangular lattice systems which have been studied.  The presence of the
double peak specific heat,  the absence of any magnetic or structural
transitions, and the gradual nature of the freezing, are all unusual.
It is therefore interesting to ask what makes \NiGaS\ special?

Very likely there are several features of the material which conspire to
produce this physics.  First, the unusual exchange paths, which are a
result of the Ni-S geometry, are critical for weakening the
nearest-neighbor exchange.  This allows for relatively large
third-neighbor exchange and biquadratic coupling.  The latter can
probably understood as arising because although the microscopic overlaps
of Ni d-orbitals with the S p-states are not small, there is an accident
of bond angles which is responsible for the small $J_1$ exchange.  As a
result, fluctuations of this angle lead to a large response, and
integrating the corresponding phonons out  naturally leads to
substantial nearest-neighbor biquadratic coupling of the sign postulated
here.  The large $J_3$ exchange leads to the incommensurate
correlations, which in turn we have argued in
Sec.~\ref{sec:imry-ma-argument} are responsible for the lack of magnetic
order and spin freezing.  The large $K$ is responsible for the
quadrupolar correlations and hence the double peak specific heat.

Also crucial to the analysis was the two dimensionality of the system.
In general, there is also some interlayer coupling.  But the very large
separation between Ni layers in \NiGaS\ makes this exceedingly weak.
The nearly complete absence of correlations between layers has indeed
been verified in neutron
scattering.\cite{nakatsuji-07}
Substantial interlayer coupling could lead to sharp magnetic and/or
quadrupolar transitions, which are not observed.  

Finally, a high degree of magnetic isotropy is required to justify the
Heisenberg treatment of the system.  Sufficient magnetic anisotropy
would render the material XY-like, leading to Kosterlitz-Thouless type
physics, and also, if strong enough, further smoothing the quadrupolar
ordering crossover (since explicit magnetic anisotropy acts like a
symmetry-breaking field to the quadrupolar order parameter).  Due to the
closed shell configuration of the Ni$^{2+}$ ions, the single-ion
anisotropy is indeed expected to be tiny.  Indeed, very small
anisotropies have been inferred from ESR and NQR
experiments.\cite{yamaguchi:180404,takeya2008sda}

\subsection{Relation to prior work}
\label{sec:relation-prior-work}

Prior theoretical work\cite{tsunetsugu-06,senthil-06,laeuchli-06}
pointed out the possibility of spin-nematic phases on the triangular
lattice, and suggested these as candidate ground states for \NiGaS.
Though this is clearly not the case, the present paper builds upon these
works and preserves quadrupolar correlations as an essential ingredient.

More recently, there have been two Monte Carlo investigations of
classical O(3) spin models related to ours.  Tamura and Kawashima
studied the model with first and third neighbor Heisenberg exchange, but
no biquadratic coupling.\cite{JPSJ.77.103002} They observed a first
order transition into a low temperature bond ordered phase with broken
$C_3$ symmetry, similar to what occurs in our model.  Kawamura and
Yamamoto studied the model with nearest-neighbor Heisenberg and
biquadratic exchange\cite{kawamura2007vit} (similar to what we discuss
in Sec.~\ref{sec:nn-model}, but crucially different in that they use
O(3) spins rather than the sSU(3) approach).  They indeed observe two
temperature scales, with the upper corresponding to quadrupolar
correlations and the lower to spin correlations.  However, because of
the classical spin treatment, the nature of these correlations is
qualitatively different from what occurs for spin $S=1$ (and what we
observe here).  In particular, in their analysis the quadrupolar correlations 
are of {\sl easy axis} type rather than {\sl easy plane}, and the magnetic
correlations are thus {\sl collinear} rather than spiral.

\subsection{Outlook}
\label{sec:outlook}

\subsubsection{Theoretical issues}
\label{sec:theoretical-issues}

This work raises a number of theoretical problems which might be
addressed in the future.  Our sSU(3) treatment captures the essential
quantum nature of the phases, but does not properly treat the quantum
critical point (QCP) itself.  The field theory of the QCP is
interesting, insofar as it may be important to consider the coupling of
the gapless ``quadrupole waves'' (Goldstone modes of the spin-nematic
order) to the magnetic order parameter.  In turn, we may expect the
universal properties of this QCP to govern the physical behavior in the
temperature range between the two specific heat peaks.  Indeed, even
classically, the {\sl spin dynamics} in this temperature regime was not
addressed here, and is important to understand for comparison to a
considerable amount of NQR data.  

It would be highly desirable to improve upon the sSU(3) approximation,
to bring it into better quantitative agreement with exact diagonalization 
results at $T=0$.  At $T=0$, one possible approach would be to adopt
a description in terms of ``tensor product states''.  The direct product 
variational form to which the sSU(3) approximation reduces at $T=0$ 
can then be viewed as the simplest, zero-entanglement type of such
a tensor product state.  
Increasing the size of the tensor, and thereby introducing some 
entanglement, would systematically improve the ground state wavefunctions.  
Generalizing the sSU(3) approximation to extend the tensor product state 
description beyond zero temperature is an interesting theoretical challenge.

\subsubsection{Generality of the results}
\label{sec:generality-results}

Though much of the paper was devoted to a Monte Carlo study of the
$J_1-J_3-K$ model on the triangular lattice within the sSU(3)
approximation, the arguments presented in Secs.~\ref{sec:symm-order-param}
and \ref{sec:imry-ma-argument} are quite general.
In particular, the emergence of two temperature scales near a
quadrupolar to magnetic QPT, and the sensitivity of two-dimensional
incommensurate spiral states to non-magnetic impurities should both
apply rather broadly to many magnetic materials.  It should be
interesting to seek applications of these ideas amongst other frustrated
magnetic materials.

\section{Acknowledgments}
This work was supported by the DOE through Basic Energy Sciences grant
DE-FG02-08ER46524. LB's research facilities at the KITP
were supported by the National Science Foundation grant NSF PHY-0551164.

\appendix 

\section*{Appendix: Definition and Implementation of Measurements}
\label{appendix:measurements}

In the following we give a detailed account of how to relate measurements of physical
observables in the original spin models \eqref{eq:1} and \eqref{eq:j1k-Hamiltonian} 
to expressions in terms of the three-component complex vectors of our semiclassical 
sSU(3) approximation.

We implemented measurements of the spin structure factor $\mathcal{S}_{{\bf k}}$ by making use of the identification
\[
{\bf S}_i =  -i {\bf b}^*_i \times {\bf b}_i \ .
\]
We then computed the structure factor by taking
\begin{eqnarray*}
{\bf A}_{{\bf k}} & = & \sum_i e^{i\,{\bf k}\cdot{\bf r}_i} {\bf S}_i \,, \\ 
\mathcal{S}_{{\bf k}} & = & \avg{|{\bf A}_{{\bf k}}|^2} \ .
\end{eqnarray*}

Similarly, we computed the quadrupolar structure factor $\mathcal{Q}_{{\bf k}}$ by measuring the quadrupolar order parameter defined by
\[
Q^{\mu\nu}_i  = \frac{1}{3} \delta^{\mu\nu} - \frac{1}{2} ( b^*_{i\mu} b_{i\nu} + b^*_{i\nu} b_{i\mu} )
\]
and then computing
\begin{eqnarray*}
B^{\mu\nu}_{{\bf k}} & = & \sum_i e^{i\,{\bf k}\cdot{\bf r}_i} Q^{\mu\nu}_i \,, \\ 
\mathcal{Q}_{{\bf k}} & = & \sum_{\mu\nu} \avg{|B^{\mu\nu}_{{\bf k}}|^2} \ .
\end{eqnarray*}

Having obtained the spin and quadrupolar structure factors, the spin and quadrupolar correlation lengths for 
ordering with wavenumber $k$ were obtained by assuming the Ornstein-Zernike (mean-field) form for the structure 
factor peaks. One then obtains the expressions
\begin{eqnarray*}
\xi^S_{k} & = & \frac{1}{2 \delta} \left( \sqrt{\frac{\avg{\mathcal{S}_{k}}}{\avg{\mathcal{S}_{k+\delta}}} - 1 } +  \sqrt{\frac{\avg{\mathcal{S}_{k}}}{\avg{\mathcal{S}_{k-\delta}}} - 1 } \right ) \,, \\
\xi^Q_{k} & = & \frac{1}{2 \delta} \left( \sqrt{\frac{\avg{\mathcal{Q}_{k}}}{\avg{\mathcal{Q}_{k+\delta}}} - 1 } +  \sqrt{\frac{\avg{\mathcal{Q}_{k}}}{\avg{\mathcal{Q}_{k-\delta}}} - 1 } \right ) \ .
\end{eqnarray*}
Here $\avg{\mathcal{S}_p}$ stands for an average of the structure factor over wavevectors of magnitude $p$ oriented along 
each of the six triangular lattice bond directions. 
The shift $\delta$ can be any small number for the case of the infinite system, but for the triangular lattice of linear dimension $L$, 
we took $\delta = 4\pi/L$ since the structure factors are well defined only at the discrete Born-von Karman momenta. 
Finally, since we assume above that the structure factors are mean-field like, the results for the correlation lengths can only be trusted above the 
respective temperature scales at which each correlation length grows to be large.

\bibliography{NiGaS_5_26}

\end{document}